\documentclass[sigplan,screen]{acmart}
\usepackage{xcolor}
\usepackage{listings}

\setcopyright{cc}
\setcctype{by}
\acmDOI{10.1145/3759429.3762632}
\acmYear{2025}
\copyrightyear{2025}
\acmISBN{979-8-4007-2151-9/25/10}
\acmConference[Onward! '25]{Proceedings of the 2025 ACM SIGPLAN International Symposium on New Ideas, New Paradigms, and Reflections on Programming and Software}{October 12--18, 2025}{Singapore, Singapore}
\acmBooktitle{Proceedings of the 2025 ACM SIGPLAN International Symposium on New Ideas, New Paradigms, and Reflections on Programming and Software (Onward! '25), October 12--18, 2025, Singapore, Singapore}
\received{2025-04-24}
\received[accepted]{2025-08-11}

\begin{document}

\title{Let’s Take Esoteric Programming Languages Seriously}

\author{Jeremy Singer}
\orcid{0000-0001-9462-6802}
\affiliation{%
  \institution{University of Glasgow}
  \city{Glasgow}
  \country{United Kingdom}
}
\email{jeremy.singer@glasgow.ac.uk}

\author{Steve Draper}
\orcid{0000-0002-9564-9317}
\affiliation{%
  \institution{University of Glasgow}
  \city{Glasgow}
  \country{United Kingdom}
}
\email{steve.draper@glasgow.ac.uk}

% General Lessons from Esoteric Languages
% The Next 700 Esoteric Programming Languages

\begin{abstract}
  Esoteric programming languages are challenging to learn, but their unusual features and constraints may serve to improve programming ability. From languages designed to be intentionally obtuse (e.g.\ INTERCAL) to others targeting artistic expression (e.g.\ Piet) or exploring the nature of computation (e.g.\ Fractan), there is rich variety in the realm of esoteric programming languages. This essay examines the counterintuitive
  appeal of esoteric languages and seeks to analyse reasons for this
  popularity.
  We will explore why people are attracted to esoteric languages in terms of (a)~program comprehension and construction, as well as (b)~language design and implementation.
  Our assertion is that esoteric languages can improve general PL awareness, at the same time as enabling the esoteric programmer to impress their peers with obscure knowledge. We will also consider pedagogic principles and the use of AI, in relation to esoteric languages.
  Emerging from the specific discussion, we identify a general set of `good' reasons for designing new programming languages. It may not be possible for anyone to be exhaustive on this topic, and it is certain we have not achieved that goal here.   However we believe
  our most important contribution is to draw more attention to
  the varied and often implicit motivations involved in programming language design.
  % how esoteric languages push the boundaries of software development.

\end{abstract}



\maketitle

%-------------------------------
% Section : Introduction
%-------------------------------
\section{Introduction}
\label{sec:intro}

Despite their name, \emph{esoteric} programming languages are surprisingly popular
in the general Computer Science community.
This essay is based on a presentation we have given several times to audiences of academic Computer Scientists interested in programming language design; on each occasion
we have encountered keen advocates of esoteric languages like
Whitespace and Brainf***. However we have also
detected a sense of shame---esoteric language usage is an unexpectedly common
`guilty secret'.
In this work, we seek to analyse potential reasons
for these attitudes to esoteric languages. We will argue that Computer Scientists need to take esoteric languages seriously.

Let's start with some definitions:

An \emph{esoteric programming language} (hereafter contracted to \emph{esolang}, following general convention) is `a programming language designed to experiment with weird ideas, to be hard to program in, or as a joke, rather than for practical use'.\footnote{\url{https://esolangs.org/wiki/Esoteric_programming_language}}
Temkin \cite{temkin2017language} characterizes an esoteric programming language
as one that is `intentionally ununsable, uncomputable or conceptual'.
Another definition\footnote{\url{https://esoteric.codes/blog/esolangs-as-an-experiential-practice}} is that `esolangs are a class of languages made for reasons other than practical use'.

More generally, the adjective \emph{esoteric} relates to
philosophical doctrines and modes of speech. It refers to concepts or materials that are appreciated only by an `inner circle' of advanced persons, known as initiates.

And, if a definition were needed, a \emph{programming language}
is a systematized notation for precise expression of intended computational behaviour.
The ACM encyclopedia definition of `programming languages' \cite{sammet2003programming} describes them as:
`both tools for directing the operation of a computer and tools for organizing and expressing solutions to problems'.

The claim of this essay is that esoteric languages might be inherently
useful in some significant ways, and so they deserve to be studied in a serious manner. The first question we seek to address is:
\textit{What is the appeal of esolangs, in general?}
Our discussion will cover the following four high-level reasons for their
popularity:
\begin{enumerate}
\item A sense of fun or playfulness
\item Recalling a lost `golden age’ of coding
\item Sense of cultic initiation
\item Artistic value
\end{enumerate}

The second question we will address is:
\textit{What is the appeal of designing esolangs, in particular?}
%April fools
We will explore the reasons of parody and comedy, before moving on to considering the skills of 
working within very tight self-imposed constraints, and the
attraction of creating a programming language `of one's own',
as Virginia Woolf might have said.

Taking inspiration from Wing's conceptual framework of
\emph{Computational Thinking}
\cite{wing2006computational},
we will ask: \textit{what do programmers actually do?}
We will explore how this characteristic behaviour is manifested in
the medium of esolangs.
Whereas Wing's claim is that practitioners in other disciplines can benefit from
a grasp of computing principles, our study of esoteric languages leads us to
believe that computer programmers can benefit from a wider appreciation of
other disciplines.
This is Gordon's argument in his Onward! essay from last year \cite{gordon2024linguistics}
which
takes the complementary stance to Wing, encouraging computer programmers to appreciate
natural language and linguistics.

In considering pedagogic principles, we will ask
\textit{Why is it beneficial to expose learners to esoteric languages?}
We will explore this question through the lens of variation theory, \emph{inter alia}.

%% There are significant problems with programming language design,
%% issues that have been conveniently ignored by
%% designers for decades - highlighted by the nature of esolangs. Mention Gordon and linguistics of programming,
%% Also PPIG seminar from 2014 - check.

Although esolangs may seem trivial, they deserve our attention since they can change the way we think about programming and language design. As Perlis \cite{perlis1982epigrams} puts it:
\begin{quotation}
  Epigram 19: A language that doesn’t affect the way you think about programming, is not worth knowing.
\end{quotation}

%-------------------------------
% Section : Background
%-------------------------------
\section{Background}
\label{sec:background}

In this section, we briefly survey the field of esolangs.
We commence with the prototypical esolang, INTERCAL, in
Section \ref{sec:bg:intercal}.
After this, for the sake of convenience we cluster the languages into
broadly related sets, considering
esolangs that have a natural language syntax (Section \ref{sec:bg:nl}), those with non-textual syntax (Section \ref{sec:bg:exosyn}) and
those with an exotic computational model (Section \ref{sec:bg:exomod}).
We acknowledge that these characteristic properties are not mutually exclusive. Further, we acknowledge that other esolangs may not fall into any of these categories, but we hope this is sufficient for an imprecise `first cut' at surveying the field.

One key observation is that almost all esolangs have
low-level semantics, in terms of primitive arithmetic operations on integer values, limited input/output facilities and
primitive control flow.
Higher-level facilities such as abstract data types, modularity
and concurrency features are almost entirely absent.

Another general observation is that the surface-level, concrete
syntax of all esolang programs is highly unconventional. Typical esolang code does not look like mainstream code, indeed, it may not
look like code at all \cite{cox2013speaking}.

\subsection{INTERCAL}
\label{sec:bg:intercal}

We first discuss INTERCAL, since this is the archetype of
esolangs. 
To the best of our knowledge, its language reference manual \cite{intercalman}
makes the first mention of `esoteric' in relation to
programming languages of this nature.
\begin{quotation}
The examples of INTERCAL programming which have appeared in the preceding sections of this manual
have probably seemed highly esoteric to the reader unfamiliar with the language. With the aim of making
them more so, we present here a description of INTERCAL.
\end{quotation}

Like many subsequent esolangs, INTERCAL was devised by
university students as a playful, humorous artifact. It has become
`the longest-running and most convoluted joke in the history of programming language design' \cite{raymond2010risks}.
Although the language initially had a minimal compiler and
negligible source code corpus, 
it caught the attention of a small group of dedicated enthusiasts
including high profile Computer Scientists like
Raymond \cite{raymondintercal} and Knuth \cite{knuth2011tpk} who helped to popularize it.

INTERCAL is not a well-formed acronym or abbreviation:
confusingly\footnote{The contraction is `for obvious reasons', according to the language manual \cite{intercalman}.} it stands for
`Compiler Language With No Pronounceable Acronym'.
The language was invented in 1972 by two Princeton students
as a parody of the verbose imperative languages
of the day.

%% Paper on intercal - \url{http://www.catb.org/esr/intercal/paper.html} (cite properly)

INTERCAL has unusual lexical elements, such as
the use of the characters \textcent\ and \textdollar\ as operators. There
are also unconventional binary operations, such as bitwise
interleavings of values. For these reasons among others,
fairly standard arithmetic operations become complex and convoluted
in INTERCAL.

One of the (many) highly original features of INTERCAL is
the \texttt{PLEASE} command modifier.
The compiler requires occasional politeness, however excessive
\texttt{PLEASE}s cause a compilation error.
In a remarkable back-to-the-future moment, recent ChatGPT
users have demonstrated similar politeness with AI prompts
\cite{please}.

Bratishenko \cite{techno} identifies the
philosophy of INTERCAL as
`technomasochism', since programmers
working with this esolang seem to derive
pleasure from their painful software development
experience.

% five operations - bitlevel.

% Silly nomenclature.

\begin{figure}
  \begin{center}
    \includegraphics[width=1.0\columnwidth]{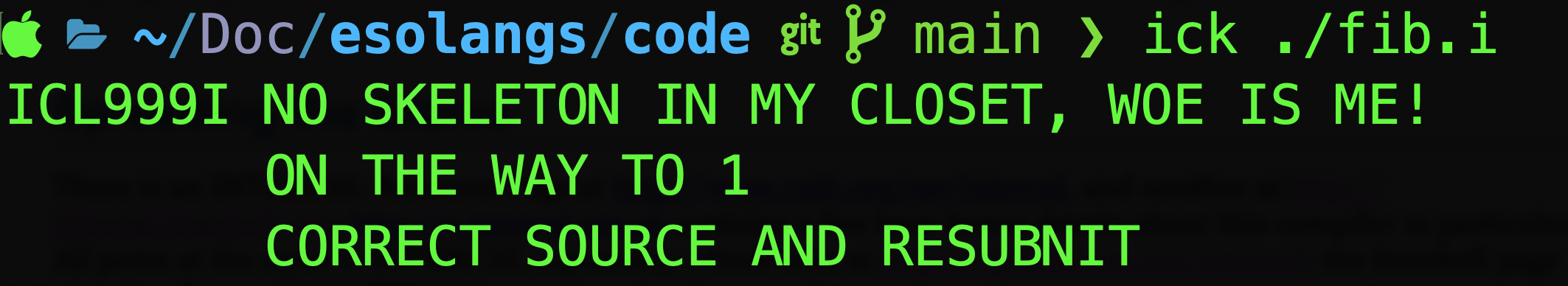}
  \end{center}
  \caption{\label{fig:bg:ick}Typical INTERCAL compiler error message}
\end{figure}

Error messages are highly unreadable, as Figure \ref{fig:bg:ick}
illustrates, when the \texttt{ick} INTERCAL compiler is unable
to parse a program correctly.

The C language version of the INTERCAL compiler improved
its reach \cite{raymondintercal}.
% Also look at Eric Raymond's resources, e.g.\
%\url{https://gitlab.com/esr/intercal}
Now, INTERCAL is available on many open-source
package management systems. For instance, you can
\texttt{brew install intercal} on macOS or
alternatively,
\texttt{apt install intercal} on Ubuntu.

We consider INTERCAL to be an archetypal esolang since
it demonstrates foundational principles that  define the genre, i.e.\ syntactic absurdity and a satirical stance toward conventional
programming language design.
Later esolangs share these playful characteristics.
INTERCAL sparked amusement and curiosity among the broader software developer community, which continues to engage with esolangs
more generally.

\subsection{Natural Language-like Programming Languages}
\label{sec:bg:nl}

Programming in natural language was an early ambition for
language developers. Initial high-level languages like COBOL reflect this
drive, to some extent. Perhaps the most `natural' programming
language today is Inform7, used for interactive text adventure
games \cite{nelson2006natural}.

However, a large number of esolangs have program syntax that
appears to be a structured form of natural language artifacts.
Table \ref{tab:natlangs} presents a few instances of this category of esolang.

\begin{table*}
  \caption{\label{tab:natlangs}Esolangs based on structured natural language patterns}
  \begin{tabular}{l|p{4cm}|p{3cm}|p{3cm}}  \hline
    Esolang & Structure & Example assignment (x=2)& Example output (print x)\\ \hline
    Shakespeare & Elizabethan theatrical script & Thou art a red rose & Open thy heart\\
    Rockstar & 1980s song lyrics & Let your love be 2 & Shout your love\\
    Chef & Cooking recipe & 2g Chocolate & Serves 1\\
    LOLCODE & Internet chat & I HAS A x R 2 & VISIBLE x\\ \hline
  \end{tabular}
\end{table*}

For such languages, although the syntax is natural, the fixed structure
of the artifact means that programs can be parsed
in a straightforward manner.
The underlying computational model is generally reasonable, involving integers, simple arithmetic operations, assignments to local variables, conditional tests and basic imperative control flow.

\begin{figure}
  \begin{center}
    \includegraphics[width=1.0\columnwidth]{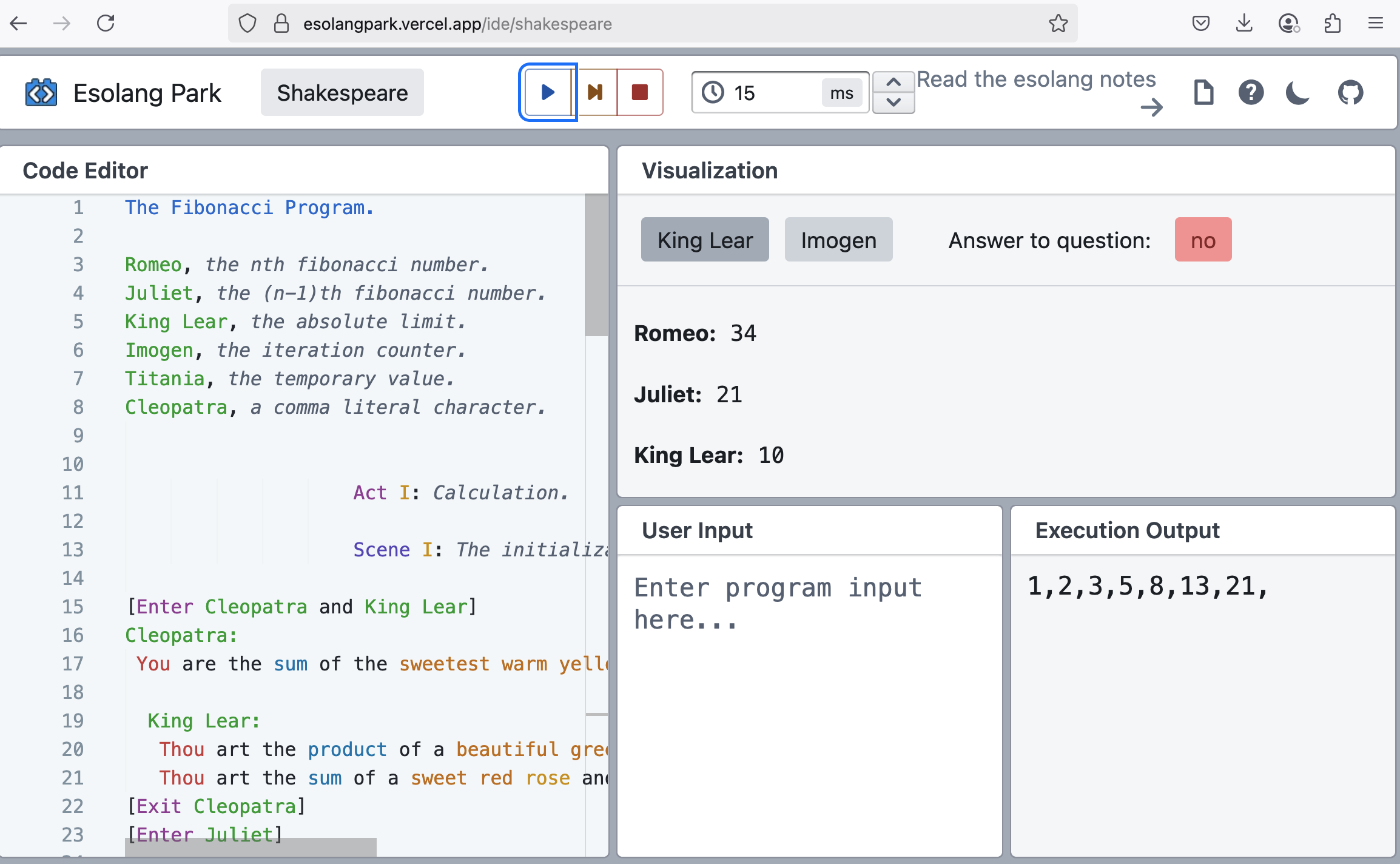}
  \end{center}
  \caption{\label{fig:bg:park}Online Interpreter for Shakespeare programs}
\end{figure}

For example, Figure \ref{fig:bg:park} shows the step-by-step
execution of a Shakespeare fibonacci calculation. Characters are individual variables storing integer values.
In Shakespeare, only two actors can be on the stage at once.
Their dialogue involves assigning values to each other.
The most complex part of Shakespeare is the representation of
integer literal constants, involving complimentary and derogatory nouns (denoting +1 and -1 respectively) with every prefixed adjective doubling a value. So `a sweet red rose' would be $2 \times 2 \times 1$, whereas a `poisonous toad' would be $2 \times -1$.

Similarly in Chef, the ingredients are used to store literal
values (with quantities) and these values are stored in variables, which are integer stacks, denoted as mixing bowls.
The various actions of cooking (pour/blend/fold) involve
computation on integer values in the bowls.

\begin{figure}
  \begin{center}
    \includegraphics[width=1.0\columnwidth]{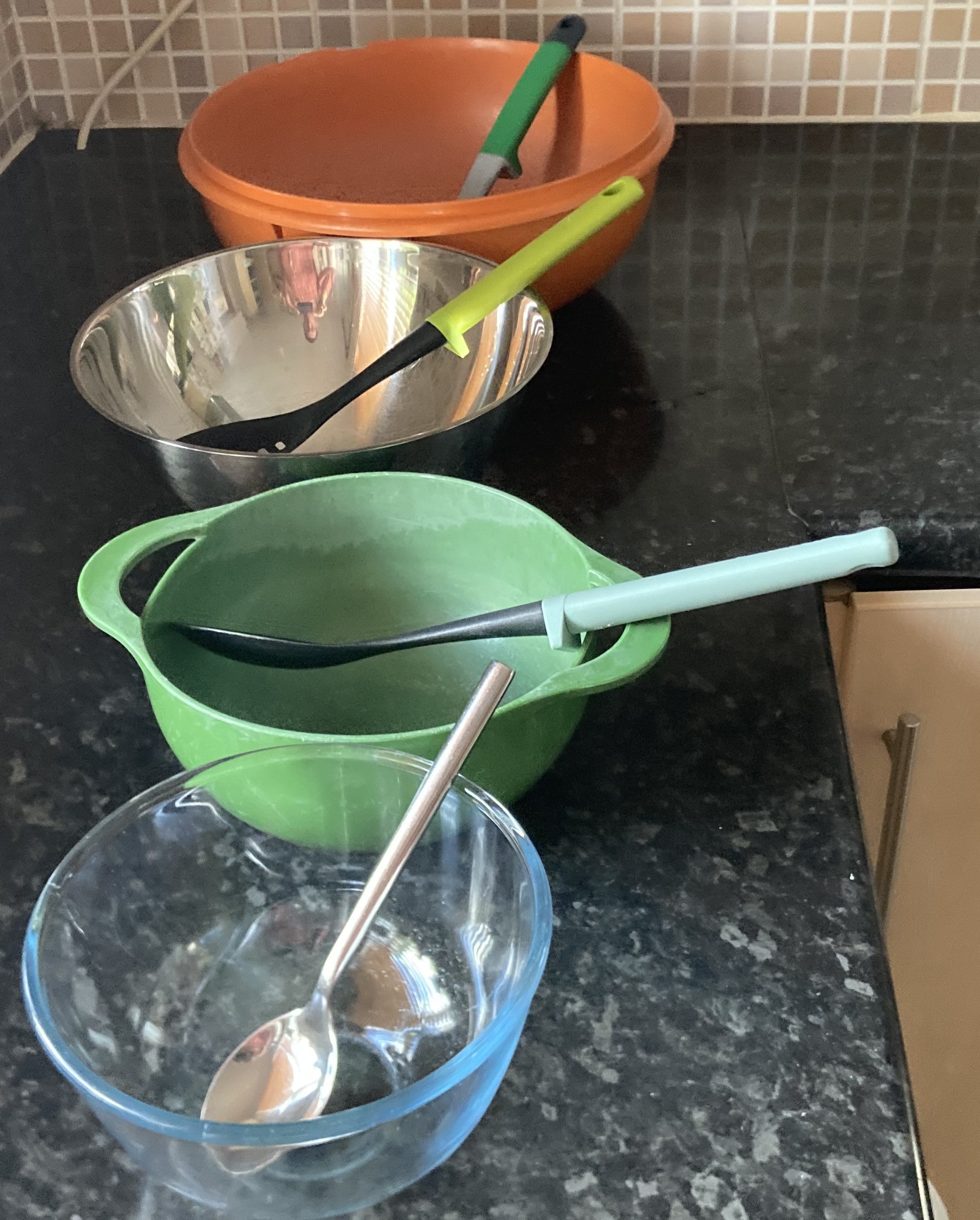}
  \end{center}
  \caption{\label{fig:bg:mixing}Set of mixing bowls, appropriate for the CHEF esolang}
\end{figure}

In general, these textual languages have primitive imperative features, often lacking facilities like dynamic memory allocation and compound data structures.

\subsection{Non-Textual Syntax}
\label{sec:bg:exosyn}

While all esolangs feature unconventional syntax,
some of them have entirely non-textual syntax.
Two such languages are Whitespace and Piet.

Both these languages feature 
relatively straightforward abstract machines.
They are 
stack based with a sensible instruction set
that has appropriate low-level operations.
In each case, the only truly unusual aspect of the
language is the syntax.

\emph{Whitespace} has only three lexemes, all of which are whitespace
characters, i.e.\ space, tab and newline.
Various combinations of these lexemes make up
instructions and literal data.

On the other hand, a \emph{Piet} program is a bitmap image.
There is a fixed palette of colours, each of which is available in three different shades.
A two-dimensional program counter, known as the
`direction pointer' traverses the image from the top-left pixel and instruction execution takes place when the direction pointer
crosses a colour boundary. The difference between colours and shades is used to perform a lookup in a two-dimensional opcode table, which has traditional stack-based operations.
The executable properties of Piet mean that a program
could be carefully recoloured or resized, without changing its semantics.

In these languages, programs deliberately don't look
like programs. For Whitespace, code is
invisible to the human eye. For Piet, code is
a colourful image. This is a kind of steganography, hiding
code in visual media.

\subsection{Unconventional Computational Model}
\label{sec:bg:exomod}

While all esolangs feature unconventional syntax,
one group of esolangs combines this with
unconventional abstract machines.

In theory, all of computation is built on
the foundation of a strange machine, i.e. the
Turing machine. As long as a
computational framework can emulate the
execution of a Turing machine, then it is Turing complete.

Mateas and Montford \cite{mateas2005box}
identify \emph{minimalist} esolangs, which have a small set
of operations and closely resemble Turing's model of
computation.
The most well known language of this kind is Brainf*** (BF).
The language has eight primitive operations
(denoted by single characters)
which mimic
a Turing machine. There is a fixed length byte array, with
operations to read and write to the current element of the array,
to move left or right to an adjacent element, to jump to a different part of the program based on the value stored in the current element, and to do character I/O.

Because of its minimal nature, BF programs can be executed with a
tiny interpreter. However most BF programs that perform `interesting' tasks are very long (like Turing machines).

Turing's assumption of an infinite-length tape
leads many programmers to assume storage space is infinite
and that execution time may be unbounded.
In contrast, BF imposes strict static limits on storage (generally the array is
30KB in size).

The Befunge esolang is somewhat similar to BF.
Again, operations have a single character.
Programs are laid out in a two-dimensional grid, rather
than BF's  one-dimensional line. Direction operators
are used to change the direction of the instruction pointer.
There is an auxiliary stack for intermediate computation.

Malbodge was intentionally designed to be the most difficult language in which to write code.
It is named after Dante's eighth circle of Hell.
The abstract machines uses ternary bits.
An instruction's precise effect depends on
its memory location. Code is automatically
self-modifying.
Effectively, a Malbodge program re-encrypts its
source code after every instruction execution.

As may be expected, there are relatively few extant
Malbodge programs and most of these are synthesized
from search-based code.

Conway's Fractan language \cite{conway1987fractran}
is a purely mathematical esolang.
A program starts with an initial integer value $N$ and
a fixed list of fractions $f_1, f_2, \ldots, f_k$.
Conway explains the abstract machine semantics as follows:
\begin{quotation}
You repeatedly multiply the integer
you have at any stage (initially $N$) by the earliest $f_i$ in
the list for which the answer is integral.
Whenever there is no such $f_i$, the game stops.
\end{quotation}
He demonstrates a program that generates the sequence of prime integers, then another that computes successive digits of $\pi$.
Conway criticizes other programming languages as having problems that `stem from
a bad choice of the underlying computational model'.

%%%%%

\subsection{Discussion}

It is commonly accepted that computer languages do not need
to be Turing complete; for instance, some domain-specific languages are not.
However, many esolangs are Turing complete.
Generally, they fall into the category of \emph{Turing tar-pits}---a phrase coined by Perlis \cite{perlis1982epigrams}.

\begin{quotation}
 Epigram 54: Beware of the Turing tar-pit in which everything is possible but nothing of interest is easy. 
\end{quotation}

Essentially, most programs are verbose and tricky to write. This was an explicit design goal of INTERCAL \cite{intercalman} and remains true for other esolangs.

Cox \cite{cox2013speaking} considers various aspects of
esolangs, presenting a philosophical and largely non-technical
appreciation.
We review further esolang literature in
Section \ref{sec:relw}.

%-------------------------------
% Section : Appeal 
%-------------------------------
\section{Appeal of Esoteric Languages}
\label{sec:appeal}

Esolangs have an
enduring sense of appeal to many computer scientists,
particularly programming language designers.
In this section, we explore the reasons for this popularity.
The characteristics we analyse have more to do with the
nature of the programmer, than of the programming language.
Although it's natural for one to reflect the other \ldots

\subsection{Playfulness}

The stereotypical hacker (in the classical sense of a computing
enthusiast) possesses a genuine 
sense of playfulness.
Raymond \cite{raymond2000hackers} makes the following observation of
this community of practice:

\begin{quotation}
  We hackers are a playful bunch; we'll hack anything, including language, if it looks like fun \ldots Deep down, we like confusing people who are stuffier and less mentally agile than we are, especially when they're bosses. There's a little bit of the mad scientist in all hackers, ready to discombobulate the world and flip authority the finger – especially if we can do it with snazzy special effects.
\end{quotation}

This playful anti-authoritarian attitude is perfectly captured by
esolangs. On the surface, programs in such languages 
appear to be confusing and counterintuitive---highly `discombobulating', as
Raymond puts it.

One way to define `play' is an activity which is defined by the activity alone, and is performed to discover the outcome,
rather than to achieve a pre-defined goal.
In this sense, the playfulness of esolangs is significant.

Really, esolangs are only appropriate for expressing `toy' programs and
simple algorithm implementations.
For the most part, esolangs feature  primitive IO, perhaps
only supporting \texttt{putchar} and \texttt{getchar} functions with no higher-level libraries.
As such, code written in esolangs is simple in terms of I/O,
probably the kind of program required in coding competitions
like the Code Olympics and the Advent of Code.
This reinforces the sense of playfulness; many people attempt Advent of Code
problems in esolangs. For instance,
a google search for `advent of code in brainf***' highlights a number of
attempts, demonstrating a non-trivial intersection between people who like  esolangs and people who enter coding competitions.

Intcode is an esolang designed explicitly for Advent of Code in 2019; really it is 
a low-level instruction set for a simple virtual machine.

%% \fixme{add quotes from the esolangs literature to back up these ideas?}

We acknowledge that it is not a new educational
idea to connect play and learning.  
% Is it time to introduce this idea more fully into CS-Education?
%Or play and learning - as in Papert.
For instance, the Logo turtle graphics language is a playful way
for school-age learners to encounter geometry, maths and programming.
Papert \cite{papert2002hard} introduces the notion of `hard fun', which
captures
`hard challenging things to do, and complex artifacts to play with'.
Further, play is experiential learning by discovery \cite{kolb2014experiential}.
To players, the \emph{activity}
is key and the \emph{outcome} is not necessarily important.
Again, these factors are evident in esolangs.

%% \fixme{ consider Idea of humour for learning.} - Steve?

Esolangs preserve a `sense of fun' in programming.
Perlis, as quoted by Abelson and Sussman \cite{abelson1996structure},
states:

\begin{quotation}
I think that it’s extraordinarily important that we in
computer science keep fun in computing. When it started out,
it was an awful lot of fun \ldots
I hope the field of
computer science never loses its sense of fun. 
\end{quotation}

\subsection{Nostalgia}
%%\subsection{Recalling a lost ‘golden age’ of coding}

It is often the case that experienced developers enjoy wearing
`the hair shirt' \cite{peytonjones2003wearing}
which involves putting themselves through
difficult experiences while developing supposedly more elegant code.

This trend originated with Turing, who invented the notion of
\emph{optimal coding}, whereby he would situate instructions at precisely
the best locations in memory to be executed with minimum delay 
in a rotary memory machine---the Pilot ACE had circulating mercury delay lines
to store instructions.
Whereas other early stored program electronic computers employed complex control circuitry to
avoid needing to make these instruction placement calculations, Turing wanted to defer
the complexity to the programmer. Campbell-Kelly records Turing's thinking
about using hardware interlocks to avoid the need for optimal coding
\cite{campbellkelly2021alan} as being:
\begin{quotation}
  \ldots much more in the American tradition of solving one's problems by means of much equipment rather than by thought.
\end{quotation}

Again, this attitude harmonizes neatly with the legendary story of Mel the Hacker from
Royal McBee, recorded in Eric Raymond's jargon file \cite{nather1983story},
describing how Mel wrote optimal code that worked perfectly on a
specific rotating memory machine due to the increment of a self-modifying instruction resulting in a jump at precisely the right moment in the program.

This idea that programming should be challenging for humans,
requiring significant mental effort, is a common one. 
The proportion of time spent thinking, relative to compiling or executing code, used to be much higher.
Of course, in the early days of computers,
compilation was much more
expensive, perhaps an overnight job.
The use of esolangs raises the level mental activity required for programming. For instance, no program was written in the Malbodge language for several years after it was designed, and the first correct program was machine-generated.

\begin{figure}
  \begin{center}
  \includegraphics[width=1.0\columnwidth]{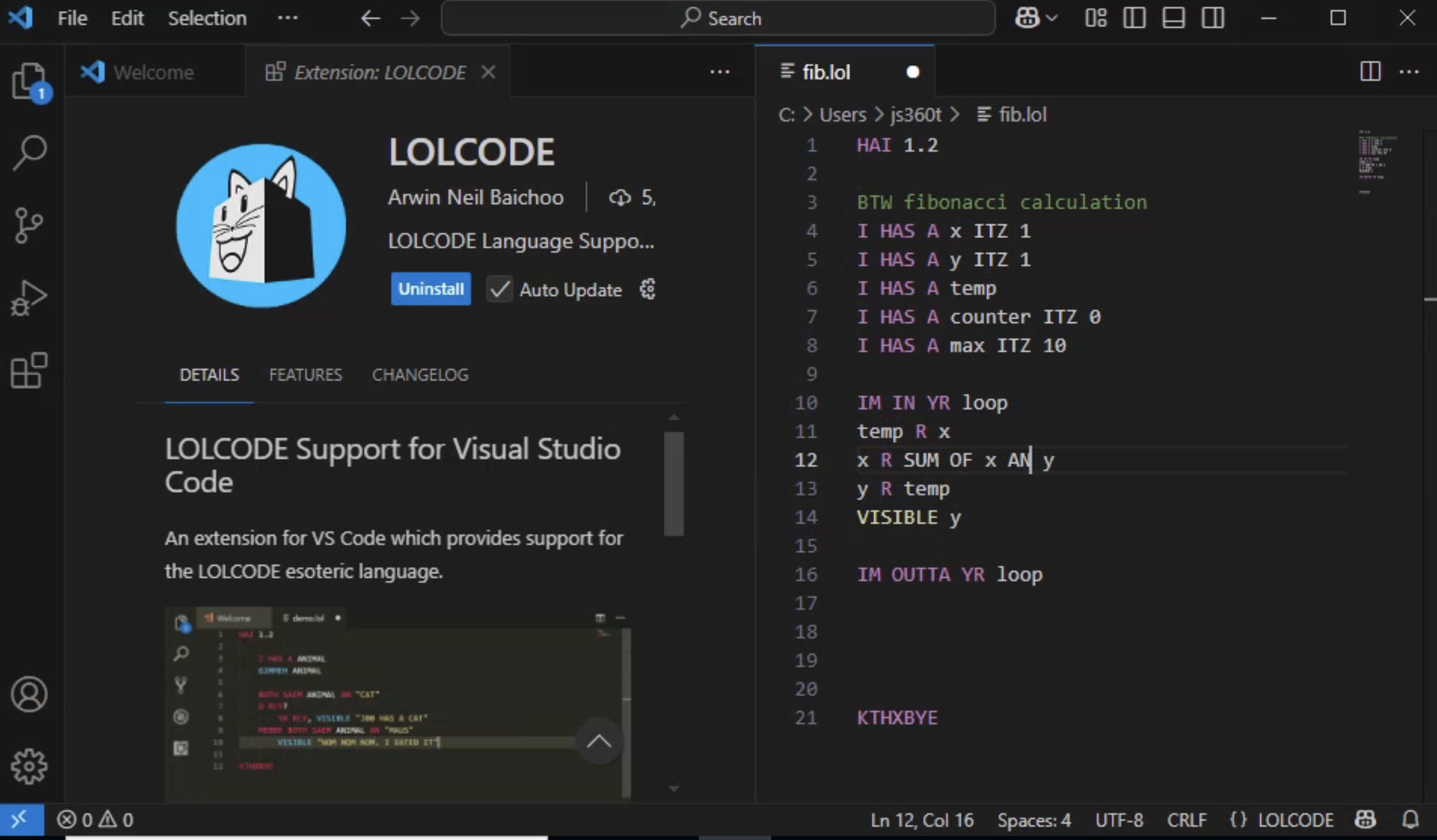}
  \end{center}
  \caption{\label{fig:vslol}Visual Studio Code with the LOLCODE plugin in operation}
\end{figure}

The key takeaway here is the hacker instinct that programming ought to be difficult, and esolangs enforce this.
By contrast, coding in mainstream languages is becoming progressively simpler.
Tooling support is another reason why coding in mainstream languages is relatively straightforward.
Many new developers use integrated development environments like Visual Studio Code, featuring syntax highlighting, autocorrection, large language model (LLM)  hints, static checking etc.
Such modern IDEs have minimal support for esolangs: LOLCODE on VS Code is
a notable exception, featuring a syntax highlighting plugin---see Figure \ref{fig:vslol}
% (see(
% \url{https://marketplace.visualstudio.com/items?itemName=arwinneil.lolcode})
for instance. Mostly developers working with esolangs just use a text editor
and terminal compilation.

Further, many mainstream managed language runtimes feature dynamic safety checks, e.g.\ for out-of-bounds memory accesses.
Even traditionally unsafe languages, like C, can be executed with address sanitizers or binary instrumentation to detect bugs.
Esolang runtimes do not have these safety checks, by default.

Another issue relates to AI-generated code.
LLMs are notoriously helpful at
generating reasonable code for mainstream languages. However
AI synthesized code is less successful for esolangs,
see Section \ref{sec:ai}.

The upshot of all of this is, for programmers hankering for the lost
`golden age' of programming, when there were no fancy development
environments or AI assistants, esoteric languages represent a
viable way to return to this time and relive this experience.
In some sense, this is a \emph{pre-Raphaelite} attitude to programming, with value placed on complexity and attention to detail.

\subsection{Sense of Belonging} % WAS cultic initiation

%% NB this section has been ChatGPT'd

Language serves as a powerful marker of community and group identity.
Fluency in a language often signifies inclusion
within a particular discourse community.
This is true not only for natural languages but also for constructed ones—consider, for instance, Creole languages or fictional tongues like Klingon.
In all these scenarios, language functions as a social bond, connecting individuals through shared forms of expression and cultural identity.

In recent programming practice,
there has been a growing emphasis on localization—developing notations that reflect a range of ethnically diverse languages as a means of promoting inclusivity
\cite{swidan2023framework}.
Some esoteric programming languages also engage with this idea of cultural identity. For example, Cree\# incorporates vocabulary from Nehiyawewin, the Plains Cree language, offering a unique intersection between programming and Indigenous linguistic heritage \cite{corbett2023cree}.
% https://pinnguaq.com/stories/cree-coding/

To the novice programmer, unfamiliar syntax and abstract concepts can feel confusing and discouraging. In contrast, experienced programmers often embrace this complexity, finding enjoyment and stimulation in the challenge. What may seem obscure or unconventional to a beginner can appear refreshingly creative and original to a seasoned practitioner \cite{cox2013speaking}.
Esoteric programming languages, in particular, foster a sense of community—not necessarily through fluency in a single language, but through a shared appreciation for playful, cryptic syntax and unconventional algorithmic expression. This shared enthusiasm hints at a kind of cultic initiation, where an ‘inner circle’ of programmers engages with code not merely as a practical tool, but as a source of amusement, aesthetic pleasure, and intellectual play.

Another key concept in esoteric practice is the idea of \emph{transmission}, which is considered a critical feature of such traditions \cite{faivre1994}.
Unlike mainstream languages, which are typically acquired via textbooks and formal instruction, esolangs are learned through more elusive means: scattered resources, word-of-mouth exchanges, and personal initiation. This difficulty in access not only shapes the learning process but also contributes to a sense of mystique, reinforcing the allure of the esoteric.

Nevertheless, the advent of the internet means it is now more straightforward
to build a community around a constructed language \cite{gobbo2005digital}.
%'The digital way to spread conlangs' - Gobbo - the internet
% really helps!

%%%

\subsection{Artistic Expression}

% Maybe put Brooks here?
In his \emph{Mythical Man Month}, Brooks
\cite{brooks1995mythical}
refers to the `sheer joy of making things' as the primary motivation for the creative acts of programmers.

Every program is an artifact, which may be viewed from different perspectives.
Lonati et al \cite{lonati2022what} enumerate six facets of programs as artifacts, dividing these
into three pairs of contrasting features, namely:
\begin{enumerate}
\item notational artefact \emph{versus} executable entity
\item human-made \emph{versus} tool
\item abstract entity \emph{versus} physical object
\end{enumerate}

Programs in esolangs emphasize:
\begin{enumerate}
\item the \emph{notational} artefact, given the weird syntax
\item the \emph{human-made}, given the ingenuity to construct the program
\item the \emph{abstract entity}, given some higher-level semantics for the abstract  machine of the particular esolang
\end{enumerate}

Knuth's literate programming \cite{knuth1984literate} conveys a deep truth which almost all programming language creators have ignored---code is written for two utterly different readers:  the compiler which acts physically on the code, and the human who reads the code to understand it.

Another perspective, acknowledged by Knuth, is that code has an \emph{aesthetic} value.
Source code may be appreciated as art, cf.\ `The Art of Computer Programming'.
In this case, the code itself is an artifact: generally (although not always) a visual object that can be enjoyed directly, by the viewer.

There are some exhibits of \emph{software as art},
for example at the Programming Language History museum.\footnote{\url{https://spectrum.ieee.org/art-of-code}}.  However it's not clear
how much of these artistic displays are source code.
%Code as art - cf.\ \url{https://www.awwsmm.com/blog/code-as-art}
%Also look at arguments on this blog post:
Esoteric languages may be considered as an art form in some circles \cite{temkin2025esolang}.

We might also mention `live coding', where the programming itself is
generating art (e.g.\ live music \cite{blackwell2005programming}).
This is usually performative art,
rather than artifact creation.
%%%(I think) - \fixme{check, and look for cites}.
More generally, usable live programming involves
fluid code editing with immediate feedback
\cite{mcdirmid2013usable}.

If artifacts can be interpreted as code, e.g.\ bitmap images as Piet programs,
or 16th Century English plays as Shakespeare programs, can we interpret
arbitrary artifacts as code? How does the real text of `Romeo and Juliet'
compute? Or is this a nonsensical question to ask? Certainly
valid Shakespeare programs respect implicit syntactic restrictions, such as allowing only two actors on-stage at any given time.

The idea of a program having multiple interpretations
was demonstrated by Worth, who wrote a Chef program that can be
interpreted
as a `Hello world' program and also interpreted
as a viable cake recipe.\footnote{\url{https://web.archive.org/web/20230501044254/https://www.mike-worth.com/2013/03/31/baking-a-hello-world-cake/}}

%%%%%

\section{Pedagogic Value}
\label{sec:pedagog}

%% Talk about how we have already hinted at this - e.g.\ with
%% earlier references to Papert.

In earlier sections, we hinted at the value of esoteric languages
with respect to learning programming.
As already discussed, Papert's notion of `hard fun' is relevant
here---esoteric languages are intriguing and may engage learners
in different ways to grapple with the challenges of programming.
We learn when we are having fun, and there is no need to be embarrassed
about this.

% Consider teaching approach - how students are to learn.
% Teaching method.
Early exposure to multiple programming languages is widely considered to be
beneficial to novice programmers \cite{denny2022novice}.
Seeing the same underlying concept (e.g.\ a counted loop)
expressed in differing contexts is
conducive to learning; this is the basis of variation theory
\cite{thune2009variation}.

Concept transfer from one programming language to another is
extremely important, helping students to enrich their notional machine models.
Identifying true- and false-carry over concepts is
highly relevant for
mainstream languages, often when they have
broadly similar syntax \cite{tshukudu2020understanding}.
For esoteric languages with unconventional syntax,
instead the deep underlying
conceptual transfers are enforced, which only have
semantic (not syntactic) correspondence.
% (examples?) (and what does Ethel call this?)

There are concrete instances of esolang usage in education.
Chef was used in the `LCC 2700 Introduction to Computational Media' course at Georgia Technology from 2005 onwards.\footnote{\url{https://www.dangermouse.net/esoteric/chef.html}}

There are many other beneficial learning outcomes when  novice programming students are exposed to esolangs.

One obvious learning outcome is the reinforcement of Dijkstra's dictum that
\texttt{GOTO} is `considered harmful' \cite{dijkstra1968letters}.
For instance, some esolangs
only have a \texttt{GOTO} style construct, such as Shakespeare's \textit{Let us proceed to Scene N}. Shakespeare code is sadly often spaghetti code.

%% Idea of the surface nature of syntax- can easily be changed.
%% - challenge conventional notions regarding program
%% source code representation.

Another positive learning outcome is the realisation that syntax is merely
the surface form of a program, and can easily be altered.
The bizarre or unexpected nature of much esolang syntax foregrounds the
fact that the surface `look-and-feel' of code is flexible.
Learners begin to understand that what really matters is the
underlying structure and semantic content behind the syntax.
This liberates novice programmers from being overly attached to any one
language (language tribalism), instead promoting a more conceptual,
language-agnostic approach to software development.

%% Idea of the limits of computation
%% Idea of the underlying computational machine model
Many esolangs pare away the conveniences of
high-level languages and expose directly the underlying computational model.
The abstract machine may be some variant of a Turing machine, or some alternative. The programmer is forced to consider how control flow, program state, and logic work at a low level.
This understanding is highly valuable, as it concretizes  what’s happening when programs execute.

There are additional learning outcomes for more advanced (i.e.\ non-novice)
students.

For instance, most esoteric languages are of the size and scale that would be suitable for a toolchain implementation as a student project.
Students would have a more rounded education if they designed an esolang
and then implemented it, using appropriate language engineering tools.
This `toe-in-the-water' experience would be relatively straightforward to
support, and would introduce learners to the issues around what makes a good
language.
% , instead of just thinking that whatever PLang sells the most licences must be the best -- the most important. 
The key critical thinking exercise here would introduce students to at least a few basic ideas on how assess  what is desirable in a programing language.

Finally, for programming educators, esolangs offer the
opportunity to 
experience similar levels of frustration as
the average novice programmer does with a mainstream language.
You might be adept at expressing quicksort in a handful of lines of Haskell,
but what happens when you try to write a sorting algorithm in an esolang?
We might be able to enhance our empathy as educators, as we relive
the agony of debugging cryptic fragments of programs we hardly understand.

So, at first glance esolangs might seem silly or impractical, but they offer serious educational value. They demonstrate that programming is not just about learning syntax or toolchains, but about reflecting critically about what we do when we write programs.

%-------------------------------
% Section : Designing new esolangs
%-------------------------------
\section{Designing New Esoteric Languages}
\label{sec:newlangs}

In this section, we discuss the design and development of
esolangs. First we examine the kinds of people who create
new esolangs (Section \ref{sec:newlangs:who}) and then
we move on to consider their reasons for creating
such languages (Section \ref{sec:newlangs:why}).

%%%%%

\subsection{Who Develops Esolangs?}
\label{sec:newlangs:who}

\subsubsection{Mostly Students?}
% students

Many esolangs were designed by Computer Scientists during
their university studies.
For instance, the INTERCAL language was created by two Princeton University
students: Woods and Lyon.
The Whitespace language was created by two undergraduates at Durham University:
Brady and Morris.
Further, the Shakespeare language was created
by students {\AA}slund and Wiberg as a 
coursework project submission in their Compilers course
at Royal Institute of Technology in Stockholm.\footnote{
\url{https://web.archive.org/web/20220721085340/http://shakespearelang.sourceforge.net/report/shakespeare/shakespeare.html}}

These instances demonstrate that the simplicity of esolangs means their implementation is tractable as a
one- or two-person programming exercise, for advanced undergraduate students.

%% \fixme{maybe Put in Steve's ILO points here?}

Following Brooks' `plan to throw one away' principle of software engineering
\cite{brooks1995mythical}, some student esolang developers have proceeded
to develop more serious, mainstream languages in later life. For example,
Brady (of Whitespace fame) later invented \emph{Idris}, a
dependently-typed functional language \cite{brady2013idris}.

Similarly, Morgan-Mar is a serial inventor of esolangs, including
Chef and Piet.\footnote{\url{https://www.dangermouse.net/esoteric/}}

\subsubsection{Mostly Males?}
% males
Similarly to common perceptions regarding mainstream languages, it seems that the majority of esolang designers are males. This is based on a cursory analysis of designer names, with some implicit cultural assumptions. A more formal analysis would be required to verify this claim. However it generally concurs with the work of Hermans on feminism in programming languages \cite{hermans2024case}, as she notes that `exclusion of women
from technology is both a cause and a result of a gendered
interpretation of what programming is and what contributions matter'.
Hermans notes that valuing `hard' and employing `difficult tools that not everyone can use'
are both male-dominant characteristics, which again, might align with
the prevailing trends of esolangs.

Beckwith and Burnett \cite{beckwith2004gender}
identify how male and female developers approach
end-user programming environments in different ways.
For instance, low self-efficacy (less common in males)
attributes failure in difficult tasks to personal lack
of ability. This effect might be exacerbated in the complex
syntactic surfaces of esolangs.

On the other hand, alternative programming paradigms, such as
\emph{programming as story telling} has been shown to appeal
more to female students \cite{kelleher2008beyond}.
It is certainly the case that some esolangs (particularly the
textual ones where programs don't look like programs)
follow the story telling pattern, in some sense.

%%%%%%%%%

\subsection{Why Develop an Esolang?}
\label{sec:newlangs:why}

\subsubsection{Comedy}
The most commonly asserted reasons
for developing an esolang involve humor.

INTERCAL is essentialy a parody of
mainstream languages from the 1970s.
The courteous \texttt{PLEASE DO} command
is a not-so-subtle criticism of the
imperative nature of FORTRAN and the
verbose syntax of COBOL.
The INTERCAL manual \cite{intercalman}
is written in a breezy, comedic style.
For example, 
the first paragraph states: `Any
resemblance of the programming language portrayed here to other programming languages, living or dead,
is purely coincidental'.

The Whitespace language was publicly announced as an April Fool's joke on the Slashdot developer news site on 1 April 2003.
The language itself was an implementation of an original idea
suggested by Stroustrup as an April Fool some five years earlier
\cite{stroustrup1998c}.

\begin{figure*}
  \begin{center}
    \includegraphics[width=2.0\columnwidth]{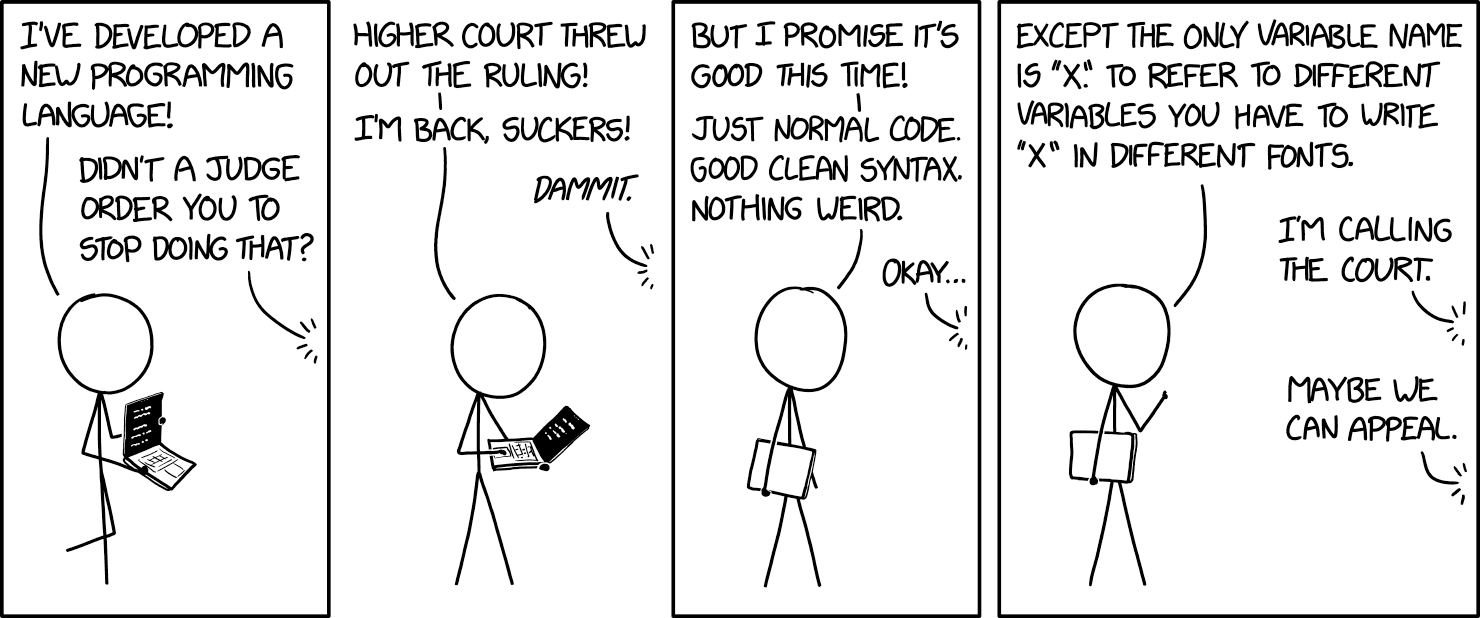}
  \end{center}
  \caption{\label{fig:xkcd}Hypothetical esolang suggestion from xkcd.com webcomic, number 2309 (CC BY-NC 2.5 Randall Munroe)}
\end{figure*}

From these two cases, we acknowledge that esolangs
are not particularly serious in general. The online web comic
\emph{xkcd} also features esolangs, see Figure \ref{fig:xkcd}.

Tiwari et al.\ \cite{tiwari2024great}
explore how humor is great motivation for software
developers. Further, Kuutila et al.\ \cite{kuutila2024what}
analyse reddit posts to characterize programming humor.
They show that `things that violate
accepted patterns in a non-threatening way are perceived to be
humorous'---attributes that clearly apply to esolang programs.

However the sheer number of esolangs and the efforts involved
in supporting them lead us to believe that there are other, deeper motivations for esolang development.

\subsubsection{Challenge}

Next we consider the skills of
working within very tight self-imposed constraints,
which are stock-in-trade for the experienced developer.

For esoteric languages, programs need to be written differently,
in an unconventional syntax and structure.
Constraints are ever-present, and in some cases programs are
practically
impossible to write, e.g.\ for the Malbodge language.
We discussed these issues earlier, in Section \ref{sec:appeal}
when we were considering the attraction of writing code in an
esolang. Similar challenges will apply to develop an esolang,
whether or not that language is self-hosting.

Developers report that challenging problems are a key
intrinsic motivation for their productivity
\cite{hall2008what}.
However again, this aspect on its own does not seem
to account for the wide variety of esolangs and the
intense development effort.

\subsubsection{Creativity}

Finally, there is the
attraction of creating a programming language `of one's own'.
Constructing an esolang permits the programmer to enjoy the
luxury of a
`break from the bland, bloated multiparadigm
languages with imperative roots'
\cite{kneusel2022strange}.
%- Temkin, in forward to
%Strange Code book.

This facility of expression, of `speaking code'
\cite{cox2013speaking}, 
is a powerful motivation for people to generate
their own, custom esolangs.
As we note in Section \ref{sec:pedagog},
creating an esolang gives excellent opportunity for students
to reflect on what make a good programming language, and to
identify tradeoffs in language design.

After all, it feels as if esolangs are probably designed
in exactly the same way as many widely used languages, i.e.\
for the convenience of the language designer rather than
of the end-user programmer.

Having considered the reasons why people develop esolangs,
which are mostly to do with creative exploration and experimentation, in the next Section we will move into a more general discussion about motivations for designing any programming language.

\section{Reasons for Developing a New Programming Language}
\label{sec:reasons}

As we study the field of esoteric programming languages, we are confronted
with a question that relates to more general programming language
philosophy:
\emph{Why does a person choose to design and develop a new programming language?} Can we identify a set of overarching motivations for
the construction of new programming languages?
Here we are using our analysis of esolangs to reflect
on wider aspects of programming language design.

In this section, we suggest five generic themes pertaining to programming language design motivation:
\begin{enumerate}
\item Expressivity
\item Efficiency
\item Education
\item Economy
\item Exploration
\end{enumerate}

We consider each of these themes in detail below,
providing example mainstream programming languages in each case.
We recognize that some languages may have multiple
design motivations.

\subsection{Expressivity}

Steele \cite{steele1999growing} beautifully illustrates the notion of
`growing a language' with increasing expressive power, which enables
succinct description of complex concepts.

One key motivation for developing a new language is the need for expressive power. Of course, computability theory tells us we could all write in Church's $\lambda$-calculus and express any computation we like, but programs would be very long and unwieldy, and programming would be complex and unenjoyable, if not intractable. More expressive notational systems allow higher-level programs to
be written. This was the original motivation for early high-level languages like Fortran, Cobol and Lisp. This is still the motivation for more modern high-level languages like Clojure or Prolog.

Jones and Bonsignour note that a programmer can write a fairly constant number of lines of code per hour, regardless
of the language used \cite{jones2011economics}.
So a more expressive language should enable programmers to be more productive. They conclude that `the overall effort associated with coding and testing are much less significant for high-level \ldots languages than for \ldots low-level languages'.

\subsection{Efficiency}

% Talk about systems implementation languages like C, Rust, Zig.
Low-level systems implementation languages like C were developed,
at least originally,
to map closely to the underlying hardware and enable efficient
code generation.
As Ritchie puts it, C is `a
simple and small language, translatable with simple and small compilers. Its types and operations are
well-grounded in those provided by real machines'
\cite{ritchie1996development}.
C has an enduring popularity \cite{kell2017some}.
%% \fixme{cite chisnall,kell,sewell etc}

More modern systems languages like Rust and Zig also
explicitly aim for efficient execution of target code
as a primary goal.

% Maybe BCPL in the olden days? - quote from Whitby Strevens book?
% High-level, but low-level.  - find a quote. Perlis?

\subsection{Education}

Another reason for developing a new language is for the purposes
of programming education.
Languages designed explicitly for novice learners include
Grace \cite{black2013seeking} although this has not been adopted
widely.
More venerable examples of educational languages include Logo and BASIC.
Scratch and Alice are instances of graphical languages
with non-traditional syntax
for school-age learners.
Hedy \cite{hermans2020hedy} is a recent programming language that is aimed at novice
learners, which has a gradually increasing syntax and semantics, to accommodate
an incremental concept inventory and learning style.

However, it is noteworthy that no mainstream industrial strength languages\footnote{One
  reviewer kindly pointed out that Pascal might be an exception here.
  }
were explicitly designed for educational use, or with respect to
pedagogical principles \cite{kolling2024principles}.
Further, initially simple industrial languages accumulate new
features throughout their lifetime, generally measured in decades  \cite{favre2005languages}.

\subsection{Economy}

Some languages are commercial imperatives---needed to satisfy economic ends.
The language that springs to mind most readily is C\#,
originally
conceived as a `direct rival' to the Java ecosystem, which seemed to be a `deep challenge to Microsoft' \cite{syme2020early}.

\subsection{Exploration}

Some programming languages are designed primarily as an intellectual exercise,
seeking to explore a new area or evaluate a new concept.
The Haskell language \cite{hudak2007history} may well fall into this category.
In his HOPL presentation of this paper, Peyton Jones
explains how Haskell was intended to `avoid success at all costs' which means the relatively small user-base are happy for the language to be `nimble' in terms of feature updates.

\subsection{Esolang Motivations}

Earlier, in Section \ref{sec:newlangs:why}, we discussed why people are motivated to design and develop new esolangs.
In this section, we want to locate those identified motivations
within the five point hierarchy of programming language design motivation outlined above (the five `E's).

For many of the motivations, it feels like esolangs
have diametrically opposed objectives.

In terms of \emph{expressivity}, esolangs deliberately use obscure syntax
and limited constructs, making it difficult to express
many algorithms in a meaningful and elegant manner.
An esolang pares down the available operations to a
minimal set, unlike most mainstream languages.

In terms of \emph{efficiency}, while esolang semantics frequently map directly onto a low-level computational model, such a model
is generally quite distinct from mainstream computer architecture, making execution less efficient.

In terms of \emph{education}, esolangs are explicitly \emph{not} designed for novice programmers. For instance, the INTERCAL
language manual \cite{intercalman} states: `Since it is an exceedingly easy language to learn,
one might expect it would be a good language for initiating novice programmers. Perhaps surprising, than,
is the fact that it would be more likely to initiate a novice into a search for another line of work'.

In terms of \emph{economy}, esolangs are not commercially viable in any sense. No meaningful software is developed in esolangs and no esolang toolchain is available on a commercial basis. Perhaps
the only people who might benefit financially from esolangs are
book authors and publishers. For instance, MIT Press has a current esolang title available \cite{cox2013speaking} and another one forthcoming \cite{temkin2025esolang}.

Therefore, it seems that the majority of esolangs
fall into the curiosity-driven, \emph{exploratory} category.
They seek to experiment with syntax or semantics in
unconventional ways, pushing the boundaries of programming in
some sense.

%%%%%%%%%%%%%%%%%

\section{AI and Esolangs}
\label{sec:ai}

No research work is complete nowadays without a consideration
of the implications of AI on the topic. In this section we
briefly discuss how AI might be useful for esolangs.

\subsection{Esolang Code Synthesis}
\begin{figure}
  \begin{center}
    \includegraphics[width=1.0\columnwidth]{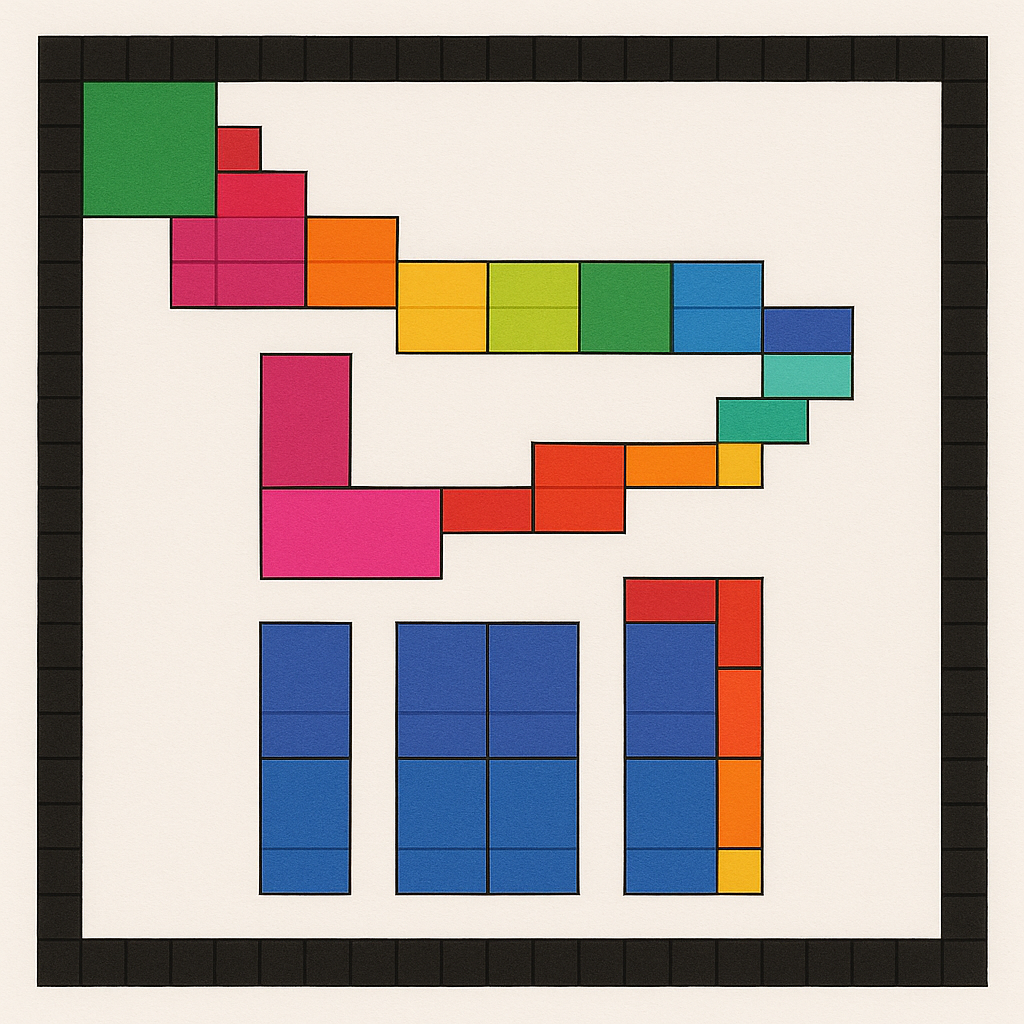}
  \end{center}
  \caption{\label{fig:ai:piet}AI-synthesized Piet Fibonacci program, which contains syntax errors}
\end{figure}

Could an AI-based large language model (LLM)
synthesize correct code to solve a specified problem
in a given esolang?
On the one hand, LLM-generated code seems to work well for mainstream languages
\cite{li2022competition,liu2023your}.
On the other hand, our initial experiments suggest that ChatGPT and Llama.cpp
do not create syntactically correct code for esolangs including
INTERCAL, Shakespeare, Whitespace and Piet.
While the textual conversation surrounding the generated code
sounds plausible, the code itself generally does not
compile and often appears to contain fundamental syntactic flaws.
For example, Figure \ref{fig:ai:piet} shows a Piet program
generated by ChatGPT, responding to our prompt to `produce a Piet
program to calculate the first 10 Fibonacci numbers'. The
resulting image cannot be parsed by the \texttt{npiet}
interpreter.

One reason why AI might be poor at synthesizing esoteric
code is the limited training set of online code examples,
in contrast to mainstream languages
like Python and JavaScript
\cite{wang2024exploring,twist2025llms}.

AI will not `render software engineering irrelevant'
\cite{kang2024tldr}. This is notably true for software
development in esolangs. The obscurity of esolangs
means they are significantly under-represented in most
LLM training corpora.

\subsection{Esolang Creation}
\begin{figure}
  \begin{center}
    \includegraphics[width=0.9\columnwidth]{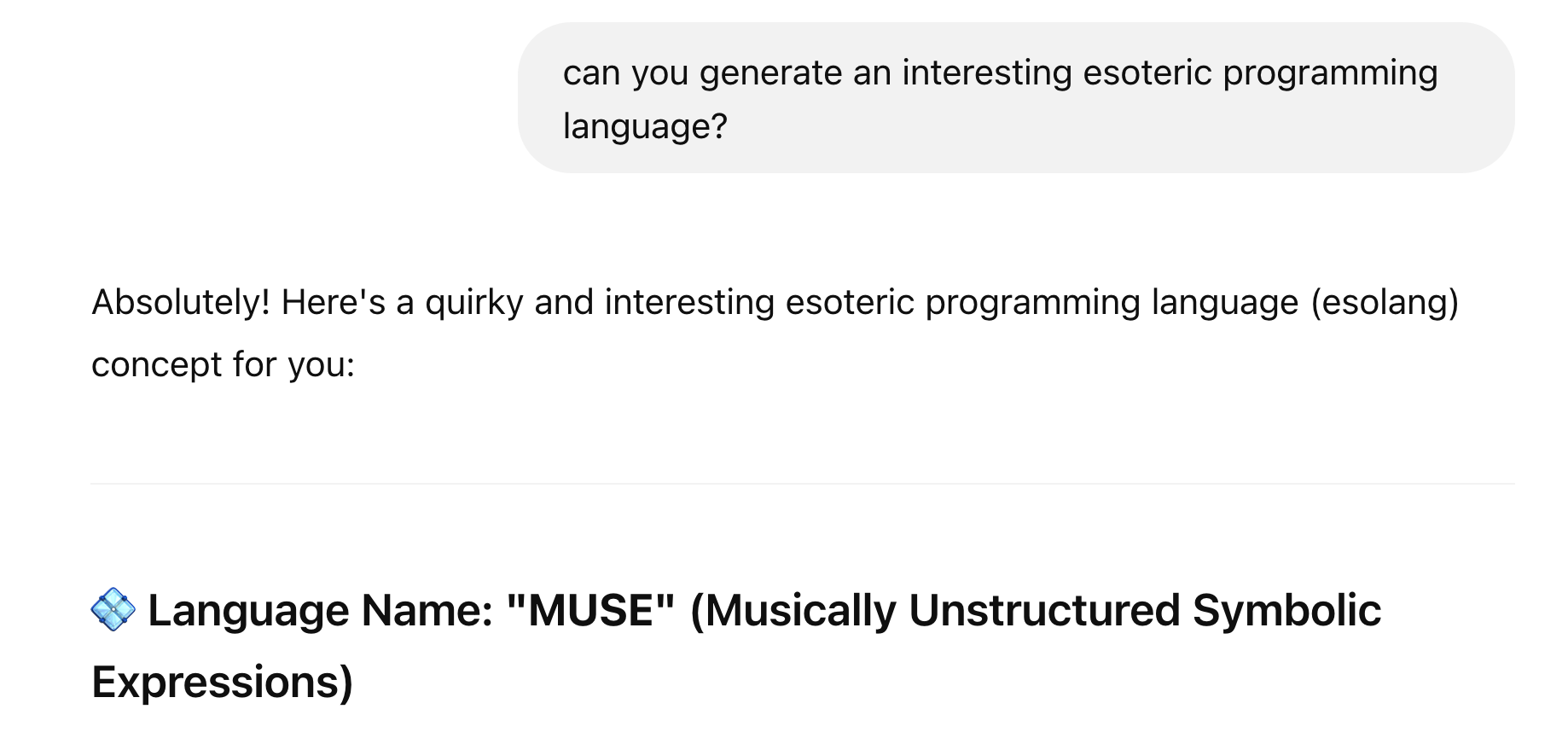}
    \includegraphics[width=0.9\columnwidth]{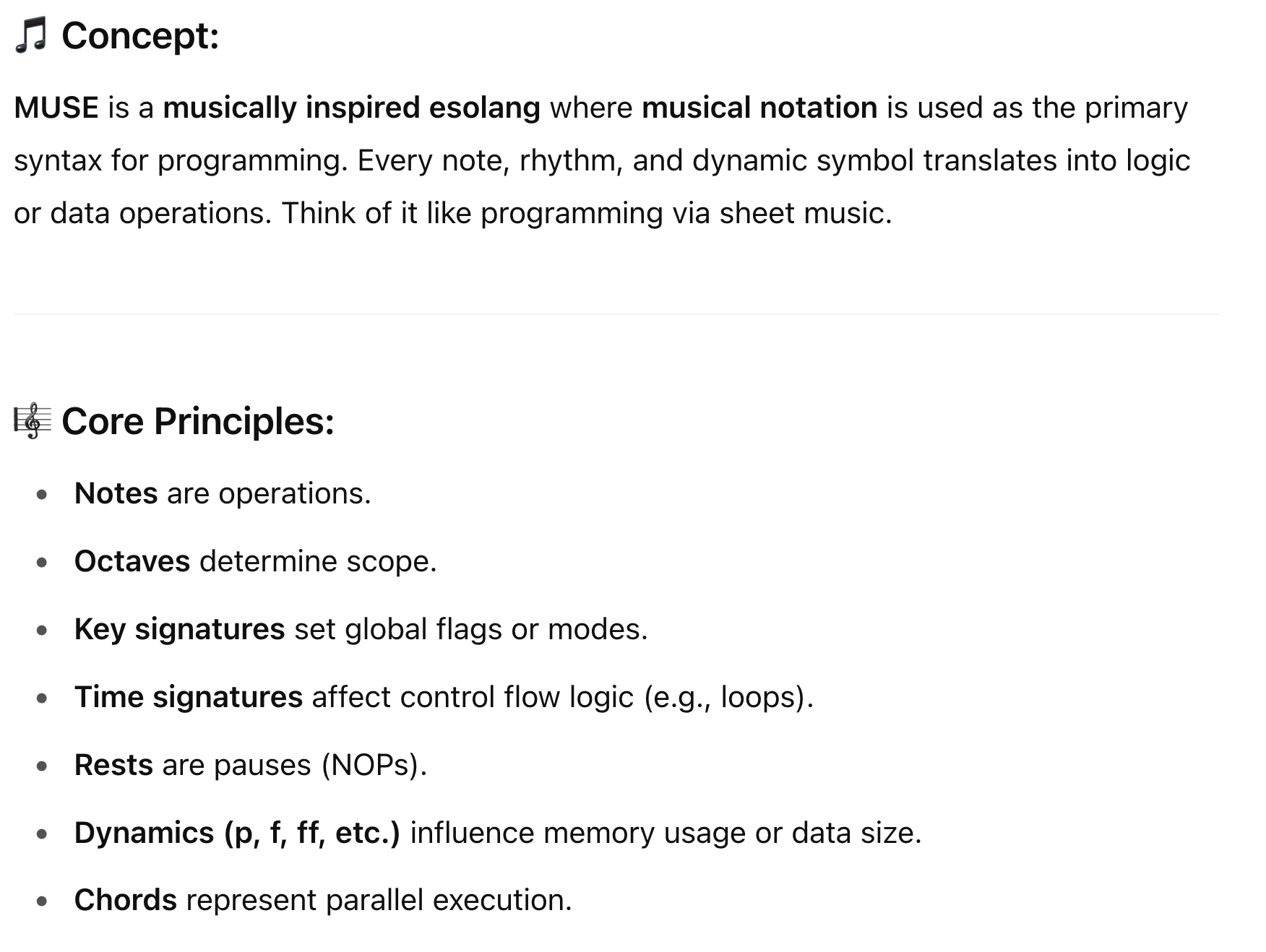}
  \end{center}
  \caption{\label{fig:ai:chatgpt}Excerpt of user interaction to develop an example esolang with ChatGPT}
\end{figure}

Could an AI model generate an interesting esolang,
either a definition or a language implementation?
We conducted some brief experiments with ChatGPT, but
the results were not particularly impressive.
Generated languages were not Turing complete, i.e.\
they did not define control flow properly.
Further, example programs were not well-formed.
Figure \ref{fig:ai:chatgpt} shows an example of a
constructed language with a syntax based on musical notation
but the language definition is both incomplete and ambiguous.
Further, this idea is highly unoriginal since a music-based
esolang is already available, named \emph{Velato}
\cite{velato}.

While there are some online tools for synthesizing constructed languages
(e.g.\ \url{https://vulgarlang.com} for fantasy novels),
these work based on user-defined parameters and employ a pseudo-random
algorithm (rather than an AI model)
which incorporates a definite notion of linguistic grammar.

%-----------------------------------
% Section : Related Work
%-----------------------------------
\section{Related Work}
\label{sec:relw}

Landin discusses the `next 700 programming languages'
\cite{landin1966next}, optimistically stating that `we
must systematize their design so that a new language is a point
chosen from a well-mapped space, rather than a laboriously
devised construction'.
Every esolang is a `laboriously devised construction', in contrast to Landin's best intentions.

Chatley et al.\ \cite{chatley2019next} envision the
`next 7000 programming languages' and discuss the reasons
for language longevity. Esolangs exhibit none of the features
required for longevity, albeit many of them are long-lived.

Mashey explores why programing languages `succeed'
\cite{mashey2004languages}. However again, the reasons
do not seem to be applicable to esolangs in general.

In the Onward! venue, there has been much discussion over
the years regarding the nature of programming languages
\cite{orchard2011four,noble2023programminglanguage}
with similar discussion at HOPL \cite{shaw2022myths}.

Steele and Gabriel 
\cite{steele50in50} present 50 languages in 50 minutes: this
live performance has taken place at a number of conference
venues over the past few years.
Their selection of languages include a number of esolangs
such as Shakespeare (Computational Drama), Befunge
(Stack Machines), and Piet (Visual Languages).
It is striking that Steele, a seasoned and successful mainstream
language designer, finds esolangs to be worthy of attention.

McIver \cite{mciver1996seven} lists seven design failures
that would impair the efficacy of novice programming languages.
Esolangs commit all seven `deadly sins', from
syntactic complexity to expectation violation.

Mateas and Montford \cite{mateas2005box}
discuss the sociological and phenomenological issues
around esolangs.
They link this community with the International Obfuscated C Code Contest
(IOCCC), which is a celebration of unconventional code expressed in a mainstream programming language (i.e.\ C) but where the syntax and semantics of the language are highly abused in order to generate code that looks unusual (some kind of ASCII art, generally) and does something unexpected when compiled (some kind of textual output, normally).
Cox \cite{cox2013speaking} explores various non-technical aspects
of esolangs in further depth.
Temkin \cite{temkin2017language} gives the most complete
review of the esolang culture and philosophy.
He seeks to explore the nature of esolangs, and the motivation
for their design and use. His findings broadly concur with ours.
His more recent study \cite{temkin2023less}
posits that esolangs enable
`personal expression and \ldots elegance within chaos'
and that they 
`challenge base assumptions of who languages are designed for and how they should be used'. Esolangs favour individual personality over practicality, and expression over comprehension.

More generally, inspired by Gordon's work relating
programming languages to linguistics
\cite{gordon2024linguistics}, we see strong linkage
between esolangs and the constructed languages
evident in modern language study---from Elvish to Esperanto
\cite{bianco2004invented}.

%-----------------------------------
% Section : Conclusion
%-----------------------------------
\section{Conclusion}
\label{sec:concl}

Esoteric languages are, by their very nature, less accessible than mainstream programming languages.
This goes against the prevailing trend, in terms of improving the accessibility of computing---whether from a usability or diversity perspective. So esolangs are implicitly contrarian, set against the direction of mainstream philosophy and advocacy.

As we have looked at esolangs, we have endeavoured to learn
something about the essence of programming, but in fact,
we appear to have uncovered something about the essence
of programmers\ldots

Precisely what? That they are a strange bunch?
It appears that esolang users and developers derive perverse pleasure from the constraints and complexities of esolangs,
engaging in the `technomasochism' identified by Bratishenko
\cite{techno}.
Perhaps this observation does not apply to all esolang
programmers, but certainly it covers a significant fraction.

Ultimately, the use and development of esolangs is a form of
\emph{freedom of expression} \cite{cox2013speaking}.
Such freedom must be preserved and valued by the community. Even if that freedom of expression allows people to express themselves in highly limited ways, the freedom to be so constrained must still be preserved.

The key point of esolangs is that they enable exploration and experimentation in a
manner that permits a
`break from the bland, bloated multiparadigm
languages with imperative roots' \cite{kneusel2022strange}.
Esolangs challenge critical thinking ability regarding both programming language design and
the nature of programming.
Finally, a restatement of Perlis' epigram is appropriate \cite{perlis1982epigrams}:
\begin{quotation}
  Epigram 19: A language that doesn’t affect the way you think about programming, is not worth knowing.
\end{quotation}
For this reason above all others, esolangs are clearly `worth knowing'.

\begin{acks}
We would like to thank the following people who gave us useful additional comments which improved this paper:  Alan Blackwell, Norman Gray, and John Levine.
This work was supported in part by UKRI/EPSRC under grant
\grantsponsor{EP/X037525/1}{EPSRC}{https://epsrc.ukri.org} \grantnum{EP/X037525/1}{M4Secure}.
\end{acks}

%%%%%%%%%%%%%%%%%%%%
\bibliographystyle{ACM-Reference-Format}
\bibliography{essay}


\begin{thebibliography}{71}

%%% the Web address.

\ifx \showCODEN    \undefined \def \showCODEN     #1{\unskip}     \fi
\ifx \showISBNx    \undefined \def \showISBNx     #1{\unskip}     \fi
\ifx \showISBNxiii \undefined \def \showISBNxiii  #1{\unskip}     \fi
\ifx \showISSN     \undefined \def \showISSN      #1{\unskip}     \fi
\ifx \showLCCN     \undefined \def \showLCCN      #1{\unskip}     \fi
\ifx \shownote     \undefined \def \shownote      #1{#1}          \fi
\ifx \showarticletitle \undefined \def \showarticletitle #1{#1}   \fi
\ifx \showURL      \undefined \def \showURL       {\relax}        \fi
% invisible to TeX
\providecommand\bibfield[2]{#2}
\providecommand\bibinfo[2]{#2}
\providecommand\natexlab[1]{#1}
\providecommand\showeprint[2][]{arXiv:#2}

\bibitem[Abelson and Sussman(1996)]%
        {abelson1996structure}
\bibfield{author}{\bibinfo{person}{Harold Abelson} {and}
  \bibinfo{person}{Gerald~Jay Sussman}.} \bibinfo{year}{1996}\natexlab{}.
\newblock \bibinfo{booktitle}{\emph{Structure and interpretation of computer
  programs}}.
\newblock \bibinfo{publisher}{MIT Press}.
\newblock


\bibitem[Beckwith and Burnett(2004)]%
        {beckwith2004gender}
\bibfield{author}{\bibinfo{person}{L. Beckwith} {and} \bibinfo{person}{M.
  Burnett}.} \bibinfo{year}{2004}\natexlab{}.
\newblock \showarticletitle{Gender: An Important Factor in End-User Programming
  Environments?}. In \bibinfo{booktitle}{\emph{2004 IEEE Symposium on Visual
  Languages - Human Centric Computing}}. \bibinfo{pages}{107--114}.
\newblock
\href{https://doi.org/10.1109/VLHCC.2004.28}{doi:\nolinkurl{10.1109/VLHCC.2004.28}}


\bibitem[Black et~al\mbox{.}(2013)]%
        {black2013seeking}
\bibfield{author}{\bibinfo{person}{Andrew~P. Black}, \bibinfo{person}{Kim~B.
  Bruce}, \bibinfo{person}{Michael Homer}, \bibinfo{person}{James Noble},
  \bibinfo{person}{Amy Ruskin}, {and} \bibinfo{person}{Richard Yannow}.}
  \bibinfo{year}{2013}\natexlab{}.
\newblock \showarticletitle{Seeking grace: a new object-oriented language for
  novices}. In \bibinfo{booktitle}{\emph{Proceeding of the 44th ACM Technical
  Symposium on Computer Science Education}}. \bibinfo{pages}{129–134}.
\newblock
\href{https://doi.org/10.1145/2445196.2445240}{doi:\nolinkurl{10.1145/2445196.2445240}}


\bibitem[Blackwell and Collins(2005)]%
        {blackwell2005programming}
\bibfield{author}{\bibinfo{person}{Alan~F Blackwell} {and}
  \bibinfo{person}{Nick Collins}.} \bibinfo{year}{2005}\natexlab{}.
\newblock \showarticletitle{The Programming Language as a Musical Instrument.}.
  In \bibinfo{booktitle}{\emph{PPIG}}. \bibinfo{pages}{11}.
\newblock
\newblock
\shownote{\url{https://www.ppig.org/files/2005-PPIG-17th-blackwell.pdf}}.


\bibitem[Brady(2013)]%
        {brady2013idris}
\bibfield{author}{\bibinfo{person}{Edwin Brady}.}
  \bibinfo{year}{2013}\natexlab{}.
\newblock \showarticletitle{Idris, a general-purpose dependently typed
  programming language: Design and implementation}.
\newblock \bibinfo{journal}{\emph{Journal of Functional Programming}}
  \bibinfo{volume}{23}, \bibinfo{number}{5} (\bibinfo{year}{2013}),
  \bibinfo{pages}{552--593}.
\newblock
\href{https://doi.org/10.1017/S095679681300018X}{doi:\nolinkurl{10.1017/S095679681300018X}}


\bibitem[Bratishenko(2009)]%
        {techno}
\bibfield{author}{\bibinfo{person}{Lev Bratishenko}.}
  \bibinfo{year}{2009}\natexlab{}.
\newblock \showarticletitle{Technomasochism: Getting spanked by {INTERCAL}}.
\newblock \bibinfo{journal}{\emph{Cabinet Magazine}}  \bibinfo{volume}{36}
  (\bibinfo{year}{2009}).
\newblock
\newblock
\shownote{\url{https://www.cabinetmagazine.org/issues/36/bratishenko.php}}.


\bibitem[Brooks~Jr(1995)]%
        {brooks1995mythical}
\bibfield{author}{\bibinfo{person}{Frederick~P Brooks~Jr}.}
  \bibinfo{year}{1995}\natexlab{}.
\newblock \bibinfo{booktitle}{\emph{The mythical man-month (anniversary ed.)}}.
\newblock \bibinfo{publisher}{Addison-Wesley}.
\newblock


\bibitem[Campbell-Kelly(2012)]%
        {campbellkelly2021alan}
\bibfield{author}{\bibinfo{person}{Martin Campbell-Kelly}.}
  \bibinfo{year}{2012}\natexlab{}.
\newblock \showarticletitle{Alan Turing's other universal machine}.
\newblock \bibinfo{journal}{\emph{Commun. ACM}} \bibinfo{volume}{55},
  \bibinfo{number}{7} (\bibinfo{date}{July} \bibinfo{year}{2012}),
  \bibinfo{pages}{31–33}.
\newblock
\href{https://doi.org/10.1145/2209249.2209277}{doi:\nolinkurl{10.1145/2209249.2209277}}


\bibitem[Chatley et~al\mbox{.}(2019)]%
        {chatley2019next}
\bibfield{author}{\bibinfo{person}{Robert Chatley}, \bibinfo{person}{Alastair
  Donaldson}, {and} \bibinfo{person}{Alan Mycroft}.}
  \bibinfo{year}{2019}\natexlab{}.
\newblock \showarticletitle{The next 7000 programming languages}.
\newblock \bibinfo{journal}{\emph{Computing and software science: State of the
  art and perspectives}} (\bibinfo{year}{2019}), \bibinfo{pages}{250--282}.
\newblock
\href{https://doi.org/10.1007/978-3-319-91908-9_1}{doi:\nolinkurl{10.1007/978-3-319-91908-9_1}}


\bibitem[Conway(1987)]%
        {conway1987fractran}
\bibfield{author}{\bibinfo{person}{John~H Conway}.}
  \bibinfo{year}{1987}\natexlab{}.
\newblock \showarticletitle{Fractran: A simple universal programming language
  for arithmetic}.
\newblock In \bibinfo{booktitle}{\emph{Open problems in Communication and
  Computation}}. \bibinfo{publisher}{Springer}, \bibinfo{pages}{4--26}.
\newblock


\bibitem[Corbett(2023)]%
        {corbett2023cree}
\bibfield{author}{\bibinfo{person}{Jon Corbett}.}
  \bibinfo{year}{2023}\natexlab{}.
\newblock \bibinfo{title}{Cree Coding}.
\newblock
\newblock
\shownote{Available from \url{https://pinnguaq.com/stories/cree-coding/}}.


\bibitem[Cox and McLean(2012)]%
        {cox2013speaking}
\bibfield{author}{\bibinfo{person}{Geoff Cox} {and} \bibinfo{person}{Alex
  McLean}.} \bibinfo{year}{2012}\natexlab{}.
\newblock \bibinfo{booktitle}{\emph{Speaking Code: Coding as Aesthetic and
  Political Expression}}.
\newblock \bibinfo{publisher}{MIT Press}.
\newblock
\showISBNx{9780262018364}
\href{https://doi.org/10.7551/mitpress/8193.001.0001}{doi:\nolinkurl{10.7551/mitpress/8193.001.0001}}


\bibitem[Denny et~al\mbox{.}(2022)]%
        {denny2022novice}
\bibfield{author}{\bibinfo{person}{Paul Denny}, \bibinfo{person}{Brett~A.
  Becker}, \bibinfo{person}{Nigel Bosch}, \bibinfo{person}{James Prather},
  \bibinfo{person}{Brent Reeves}, {and} \bibinfo{person}{Jacqueline Whalley}.}
  \bibinfo{year}{2022}\natexlab{}.
\newblock \showarticletitle{Novice Reflections During the Transition to a New
  Programming Language}. In \bibinfo{booktitle}{\emph{Proceedings of the 53rd
  ACM Technical Symposium on Computer Science Education - Volume 1}}.
  \bibinfo{pages}{948–954}.
\newblock
\href{https://doi.org/10.1145/3478431.3499314}{doi:\nolinkurl{10.1145/3478431.3499314}}


\bibitem[Dijkstra(1968)]%
        {dijkstra1968letters}
\bibfield{author}{\bibinfo{person}{Edsger~W Dijkstra}.}
  \bibinfo{year}{1968}\natexlab{}.
\newblock \showarticletitle{Go to statement considered harmful}.
\newblock \bibinfo{journal}{\emph{Commun. ACM}} \bibinfo{volume}{11},
  \bibinfo{number}{3} (\bibinfo{year}{1968}), \bibinfo{pages}{147--148}.
\newblock


\bibitem[Faivre(1994)]%
        {faivre1994}
\bibfield{author}{\bibinfo{person}{Antoine Faivre}.}
  \bibinfo{year}{1994}\natexlab{}.
\newblock \bibinfo{booktitle}{\emph{Access to Western esotericism}}.
\newblock \bibinfo{publisher}{SUNY Press}.
\newblock


\bibitem[Favre(2005)]%
        {favre2005languages}
\bibfield{author}{\bibinfo{person}{J.-M. Favre}.}
  \bibinfo{year}{2005}\natexlab{}.
\newblock \showarticletitle{Languages evolve too! Changing the software time
  scale}. In \bibinfo{booktitle}{\emph{Eighth International Workshop on
  Principles of Software Evolution}}. \bibinfo{pages}{33--42}.
\newblock
\href{https://doi.org/10.1109/IWPSE.2005.22}{doi:\nolinkurl{10.1109/IWPSE.2005.22}}


\bibitem[Gobbo et~al\mbox{.}(2005)]%
        {gobbo2005digital}
\bibfield{author}{\bibinfo{person}{Federico Gobbo} {et~al\mbox{.}}}
  \bibinfo{year}{2005}\natexlab{}.
\newblock \showarticletitle{The digital way to spread conlangs}.
\newblock \bibinfo{journal}{\emph{Language at Work: Language Learning,
  Discourse, and Translation Studies in Internet}} (\bibinfo{year}{2005}),
  \bibinfo{pages}{45--53}.
\newblock


\bibitem[Gordon(2024)]%
        {gordon2024linguistics}
\bibfield{author}{\bibinfo{person}{Colin~S. Gordon}.}
  \bibinfo{year}{2024}\natexlab{}.
\newblock \showarticletitle{The Linguistics of Programming}.
  \bibinfo{pages}{162–182}.
\newblock
\href{https://doi.org/10.1145/3689492.3689806}{doi:\nolinkurl{10.1145/3689492.3689806}}


\bibitem[Hall et~al\mbox{.}(2008)]%
        {hall2008what}
\bibfield{author}{\bibinfo{person}{Tracy Hall}, \bibinfo{person}{Helen Sharp},
  \bibinfo{person}{Sarah Beecham}, \bibinfo{person}{Nathan Baddoo}, {and}
  \bibinfo{person}{Hugh Robinson}.} \bibinfo{year}{2008}\natexlab{}.
\newblock \showarticletitle{What Do We Know about Developer Motivation?}
\newblock \bibinfo{journal}{\emph{IEEE Software}} \bibinfo{volume}{25},
  \bibinfo{number}{4} (\bibinfo{year}{2008}), \bibinfo{pages}{92--94}.
\newblock
\href{https://doi.org/10.1109/MS.2008.105}{doi:\nolinkurl{10.1109/MS.2008.105}}


\bibitem[Hermans(2020)]%
        {hermans2020hedy}
\bibfield{author}{\bibinfo{person}{Felienne Hermans}.}
  \bibinfo{year}{2020}\natexlab{}.
\newblock \showarticletitle{Hedy: A Gradual Language for Programming
  Education}. In \bibinfo{booktitle}{\emph{Proceedings of the 2020 ACM
  Conference on International Computing Education Research}}.
  \bibinfo{pages}{259–270}.
\newblock
\href{https://doi.org/10.1145/3372782.3406262}{doi:\nolinkurl{10.1145/3372782.3406262}}


\bibitem[Hermans and Schlesinger(2024)]%
        {hermans2024case}
\bibfield{author}{\bibinfo{person}{Felienne Hermans} {and} \bibinfo{person}{Ari
  Schlesinger}.} \bibinfo{year}{2024}\natexlab{}.
\newblock \showarticletitle{A Case for Feminism in Programming Language
  Design}. In \bibinfo{booktitle}{\emph{Proceedings of the 2024 ACM SIGPLAN
  International Symposium on New Ideas, New Paradigms, and Reflections on
  Programming and Software}}. \bibinfo{pages}{205–222}.
\newblock
\href{https://doi.org/10.1145/3689492.3689809}{doi:\nolinkurl{10.1145/3689492.3689809}}


\bibitem[Hudak et~al\mbox{.}(2007)]%
        {hudak2007history}
\bibfield{author}{\bibinfo{person}{Paul Hudak}, \bibinfo{person}{John Hughes},
  \bibinfo{person}{Simon Peyton~Jones}, {and} \bibinfo{person}{Philip Wadler}.}
  \bibinfo{year}{2007}\natexlab{}.
\newblock \showarticletitle{A history of Haskell: being lazy with class}. In
  \bibinfo{booktitle}{\emph{Proceedings of the Third ACM SIGPLAN Conference on
  History of Programming Languages}}. \bibinfo{pages}{12–1–12–55}.
\newblock
\href{https://doi.org/10.1145/1238844.1238856}{doi:\nolinkurl{10.1145/1238844.1238856}}


\bibitem[Jones and Bonsignour(2011)]%
        {jones2011economics}
\bibfield{author}{\bibinfo{person}{Capers Jones} {and} \bibinfo{person}{Olivier
  Bonsignour}.} \bibinfo{year}{2011}\natexlab{}.
\newblock \bibinfo{booktitle}{\emph{The Economics of Software Quality}}.
\newblock \bibinfo{publisher}{Addison-Wesley Professional}.
\newblock


\bibitem[Kang and Shaw(2024)]%
        {kang2024tldr}
\bibfield{author}{\bibinfo{person}{Eunsuk Kang} {and} \bibinfo{person}{Mary
  Shaw}.} \bibinfo{year}{2024}\natexlab{}.
\newblock \showarticletitle{tl;dr: {C}hill, y’all: {AI} Will Not Devour
  {SE}}. In \bibinfo{booktitle}{\emph{Proceedings of the 2024 ACM SIGPLAN
  International Symposium on New Ideas, New Paradigms, and Reflections on
  Programming and Software}}. \bibinfo{pages}{303–315}.
\newblock
\href{https://doi.org/10.1145/3689492.3689816}{doi:\nolinkurl{10.1145/3689492.3689816}}


\bibitem[Kell(2017)]%
        {kell2017some}
\bibfield{author}{\bibinfo{person}{Stephen Kell}.}
  \bibinfo{year}{2017}\natexlab{}.
\newblock \showarticletitle{Some were meant for {C}: the endurance of an
  unmanageable language}. In \bibinfo{booktitle}{\emph{Proceedings of the 2017
  ACM SIGPLAN International Symposium on New Ideas, New Paradigms, and
  Reflections on Programming and Software}}. \bibinfo{pages}{229–245}.
\newblock
\href{https://doi.org/10.1145/3133850.3133867}{doi:\nolinkurl{10.1145/3133850.3133867}}


\bibitem[Kelleher(2008)]%
        {kelleher2008beyond}
\bibfield{author}{\bibinfo{person}{Caitlin Kelleher}.}
  \bibinfo{year}{2008}\natexlab{}.
\newblock \showarticletitle{Using Storytelling to Introduce Girls to Computer
  Programming}.
\newblock In \bibinfo{booktitle}{\emph{Beyond Barbie and Mortal Kombat: New
  Perspectives on Gender and Gaming}}. \bibinfo{publisher}{MIT Press}.
\newblock
\href{https://doi.org/10.7551/mitpress/7477.003.0022}{doi:\nolinkurl{10.7551/mitpress/7477.003.0022}}


\bibitem[Kneusel(2022)]%
        {kneusel2022strange}
\bibfield{author}{\bibinfo{person}{Ronald Kneusel}.}
  \bibinfo{year}{2022}\natexlab{}.
\newblock \bibinfo{booktitle}{\emph{Strange Code: Esoteric Languages That Make
  Programming Fun Again}}.
\newblock \bibinfo{publisher}{No Starch Press}.
\newblock
\showISBNx{9781718502406}


\bibitem[Knuth(1984)]%
        {knuth1984literate}
\bibfield{author}{\bibinfo{person}{Donald~Ervin Knuth}.}
  \bibinfo{year}{1984}\natexlab{}.
\newblock \showarticletitle{Literate programming}.
\newblock \bibinfo{journal}{\emph{Comput. J.}} \bibinfo{volume}{27},
  \bibinfo{number}{2} (\bibinfo{year}{1984}), \bibinfo{pages}{97--111}.
\newblock


\bibitem[Knuth(2011)]%
        {knuth2011tpk}
\bibfield{author}{\bibinfo{person}{Donald~E. Knuth}.}
  \bibinfo{year}{2011}\natexlab{}.
\newblock \showarticletitle{Chapter 7: TPK in INTERCAL}.
\newblock In \bibinfo{booktitle}{\emph{Selected Papers on Fun and Games}}.
  \bibinfo{publisher}{Center for the Study of Language and Information},
  \bibinfo{address}{Stanford, California}.
\newblock
\showISBNx{978-1-57586-585-0}


\bibitem[Kolb(2014)]%
        {kolb2014experiential}
\bibfield{author}{\bibinfo{person}{David~A Kolb}.}
  \bibinfo{year}{2014}\natexlab{}.
\newblock \bibinfo{booktitle}{\emph{Experiential learning: Experience as the
  source of learning and development}}.
\newblock \bibinfo{publisher}{FT press}.
\newblock


\bibitem[K{\"o}lling(2024)]%
        {kolling2024principles}
\bibfield{author}{\bibinfo{person}{Michael K{\"o}lling}.}
  \bibinfo{year}{2024}\natexlab{}.
\newblock \showarticletitle{Principles of Educational Programming Language
  Design}.
\newblock \bibinfo{journal}{\emph{Informatics in Education-An International
  Journal}} \bibinfo{volume}{23}, \bibinfo{number}{4} (\bibinfo{year}{2024}),
  \bibinfo{pages}{823--836}.
\newblock
\href{https://doi.org/10.15388/infedu.2024.29}{doi:\nolinkurl{10.15388/infedu.2024.29}}


\bibitem[Kuutila et~al\mbox{.}(2024)]%
        {kuutila2024what}
\bibfield{author}{\bibinfo{person}{Miikka Kuutila}, \bibinfo{person}{Leevi
  Rantala}, \bibinfo{person}{Junhao Li}, \bibinfo{person}{Simo Hosio}, {and}
  \bibinfo{person}{Mika M\"{a}ntyl\"{a}}.} \bibinfo{year}{2024}\natexlab{}.
\newblock \showarticletitle{What Makes Programmers Laugh? Exploring the
  Submissions of the Subreddit r/ProgrammerHumor.}. In
  \bibinfo{booktitle}{\emph{Proceedings of the 18th ACM/IEEE International
  Symposium on Empirical Software Engineering and Measurement}}.
  \bibinfo{pages}{371–381}.
\newblock
\href{https://doi.org/10.1145/3674805.3686696}{doi:\nolinkurl{10.1145/3674805.3686696}}


\bibitem[Landin(1966)]%
        {landin1966next}
\bibfield{author}{\bibinfo{person}{Peter~J Landin}.}
  \bibinfo{year}{1966}\natexlab{}.
\newblock \showarticletitle{The next 700 programming languages}.
\newblock \bibinfo{journal}{\emph{Commun. ACM}} \bibinfo{volume}{9},
  \bibinfo{number}{3} (\bibinfo{year}{1966}), \bibinfo{pages}{157--166}.
\newblock


\bibitem[Li et~al\mbox{.}(2022)]%
        {li2022competition}
\bibfield{author}{\bibinfo{person}{Yujia Li}, \bibinfo{person}{David Choi},
  \bibinfo{person}{Junyoung Chung}, \bibinfo{person}{Nate Kushman},
  \bibinfo{person}{Julian Schrittwieser}, \bibinfo{person}{R{\'e}mi Leblond},
  \bibinfo{person}{Tom Eccles}, \bibinfo{person}{James Keeling},
  \bibinfo{person}{Felix Gimeno}, \bibinfo{person}{Agustin Dal~Lago},
  {et~al\mbox{.}}} \bibinfo{year}{2022}\natexlab{}.
\newblock \showarticletitle{Competition-level code generation with alphacode}.
\newblock \bibinfo{journal}{\emph{Science}} \bibinfo{volume}{378},
  \bibinfo{number}{6624} (\bibinfo{year}{2022}), \bibinfo{pages}{1092--1097}.
\newblock
\href{https://doi.org/10.1126/science.abq1158}{doi:\nolinkurl{10.1126/science.abq1158}}


\bibitem[Liu et~al\mbox{.}(2023)]%
        {liu2023your}
\bibfield{author}{\bibinfo{person}{Jiawei Liu}, \bibinfo{person}{Chunqiu~Steven
  Xia}, \bibinfo{person}{Yuyao Wang}, {and} \bibinfo{person}{Lingming Zhang}.}
  \bibinfo{year}{2023}\natexlab{}.
\newblock \showarticletitle{Is your code generated by chatgpt really correct?
  rigorous evaluation of large language models for code generation}.
\newblock \bibinfo{journal}{\emph{Advances in Neural Information Processing
  Systems}}  \bibinfo{volume}{36} (\bibinfo{year}{2023}),
  \bibinfo{pages}{21558--21572}.
\newblock


\bibitem[Lo~Bianco(2004)]%
        {bianco2004invented}
\bibfield{author}{\bibinfo{person}{Joseph Lo~Bianco}.}
  \bibinfo{year}{2004}\natexlab{}.
\newblock \showarticletitle{Invented languages and new worlds}.
\newblock \bibinfo{journal}{\emph{English Today}} \bibinfo{volume}{20},
  \bibinfo{number}{2} (\bibinfo{year}{2004}), \bibinfo{pages}{8–18}.
\newblock
\href{https://doi.org/10.1017/S0266078404002032}{doi:\nolinkurl{10.1017/S0266078404002032}}


\bibitem[Lonati et~al\mbox{.}(2022)]%
        {lonati2022what}
\bibfield{author}{\bibinfo{person}{Violetta Lonati}, \bibinfo{person}{Andrej
  Brodnik}, \bibinfo{person}{Tim Bell}, \bibinfo{person}{Andrew~Paul
  Csizmadia}, \bibinfo{person}{Liesbeth De~Mol}, \bibinfo{person}{Henry
  Hickman}, \bibinfo{person}{Therese Keane}, \bibinfo{person}{Claudio Mirolo},
  {and} \bibinfo{person}{Mattia Monga}.} \bibinfo{year}{2022}\natexlab{}.
\newblock \showarticletitle{What We Talk About When We Talk About Programs}. In
  \bibinfo{booktitle}{\emph{Proceedings of the 2022 Working Group Reports on
  Innovation and Technology in Computer Science Education}}.
  \bibinfo{pages}{117–164}.
\newblock
\href{https://doi.org/10.1145/3571785.3574125}{doi:\nolinkurl{10.1145/3571785.3574125}}


\bibitem[Mashey(2004)]%
        {mashey2004languages}
\bibfield{author}{\bibinfo{person}{John~R. Mashey}.}
  \bibinfo{year}{2004}\natexlab{}.
\newblock \showarticletitle{Languages, Levels, Libraries, and Longevity: New
  programming languages are born every day. Why do some succeed and some fail?}
\newblock \bibinfo{journal}{\emph{Queue}} \bibinfo{volume}{2},
  \bibinfo{number}{9} (\bibinfo{date}{Dec.} \bibinfo{year}{2004}),
  \bibinfo{pages}{32–38}.
\newblock
\href{https://doi.org/10.1145/1039511.1039532}{doi:\nolinkurl{10.1145/1039511.1039532}}


\bibitem[Mateas and Montfort(2005)]%
        {mateas2005box}
\bibfield{author}{\bibinfo{person}{Michael Mateas} {and} \bibinfo{person}{Nick
  Montfort}.} \bibinfo{year}{2005}\natexlab{}.
\newblock \showarticletitle{A Box, Darkly: Obfuscation, Weird Languages, and
  Code Aesthetics}. In \bibinfo{booktitle}{\emph{Proceedings of the 6th Digital
  Arts and Culture Conference}} (IT University of Copenhagen).
  \bibinfo{pages}{144–153}.
\newblock
\urldef\tempurl%
\url{https://nickm.com/cis/a_box_darkly.pdf}
\showURL{%
\tempurl}


\bibitem[McDirmid(2013)]%
        {mcdirmid2013usable}
\bibfield{author}{\bibinfo{person}{Sean McDirmid}.}
  \bibinfo{year}{2013}\natexlab{}.
\newblock \showarticletitle{Usable live programming}. In
  \bibinfo{booktitle}{\emph{Proceedings of the 2013 ACM International Symposium
  on New Ideas, New Paradigms, and Reflections on Programming and Software}}.
  \bibinfo{pages}{53–62}.
\newblock
\href{https://doi.org/10.1145/2509578.2509585}{doi:\nolinkurl{10.1145/2509578.2509585}}


\bibitem[McIver and Conway(1996)]%
        {mciver1996seven}
\bibfield{author}{\bibinfo{person}{L. McIver} {and} \bibinfo{person}{D.
  Conway}.} \bibinfo{year}{1996}\natexlab{}.
\newblock \showarticletitle{Seven deadly sins of introductory programming
  language design}. In \bibinfo{booktitle}{\emph{Proceedings 1996 International
  Conference Software Engineering: Education and Practice}}.
  \bibinfo{pages}{309--316}.
\newblock
\href{https://doi.org/10.1109/SEEP.1996.534015}{doi:\nolinkurl{10.1109/SEEP.1996.534015}}


\bibitem[Nather(1983)]%
        {nather1983story}
\bibfield{author}{\bibinfo{person}{Ed Nather}.}
  \bibinfo{year}{1983}\natexlab{}.
\newblock \bibinfo{title}{The Story of Mel}.
\newblock
\newblock
\shownote{Available from
  \url{http://www.catb.org/jargon/html/story-of-mel.html}}.


\bibitem[Nelson(2006)]%
        {nelson2006natural}
\bibfield{author}{\bibinfo{person}{Graham Nelson}.}
  \bibinfo{year}{2006}\natexlab{}.
\newblock \bibinfo{booktitle}{\emph{Natural Language, Semantic Analysis And
  Interactive Fiction}}.
\newblock \bibinfo{pages}{141--188}.
\newblock
\newblock
\shownote{\url{https://www.ifarchive.org/if-archive/books/IFTheoryBook.pdf}}.


\bibitem[Noble and Biddle(2023)]%
        {noble2023programminglanguage}
\bibfield{author}{\bibinfo{person}{James Noble} {and} \bibinfo{person}{Robert
  Biddle}.} \bibinfo{year}{2023}\natexlab{}.
\newblock \showarticletitle{programmingLanguage as Language}. In
  \bibinfo{booktitle}{\emph{Proceedings of the 2023 ACM SIGPLAN International
  Symposium on New Ideas, New Paradigms, and Reflections on Programming and
  Software}}. \bibinfo{pages}{191--204}.
\newblock
\href{https://doi.org/10.1145/3622758.3622885}{doi:\nolinkurl{10.1145/3622758.3622885}}


\bibitem[Orchard(2011)]%
        {orchard2011four}
\bibfield{author}{\bibinfo{person}{Dominic Orchard}.}
  \bibinfo{year}{2011}\natexlab{}.
\newblock \showarticletitle{The four {R}s of programming language design}. In
  \bibinfo{booktitle}{\emph{Proceedings of the 10th SIGPLAN symposium on New
  Ideas, New Paradigms, and Reflections on Programming and Software}}.
  \bibinfo{pages}{157--162}.
\newblock
\href{https://doi.org/10.1145/2089131.2089138}{doi:\nolinkurl{10.1145/2089131.2089138}}


\bibitem[Papert(2002)]%
        {papert2002hard}
\bibfield{author}{\bibinfo{person}{Seymour Papert}.}
  \bibinfo{year}{2002}\natexlab{}.
\newblock \bibinfo{title}{Hard Fun}.
\newblock
\newblock
\shownote{Available from \url{http://www.papert.org/articles/HardFun.html}}.


\bibitem[Perlis(1982)]%
        {perlis1982epigrams}
\bibfield{author}{\bibinfo{person}{Alan~J. Perlis}.}
  \bibinfo{year}{1982}\natexlab{}.
\newblock \showarticletitle{Special Feature: Epigrams on programming}.
\newblock \bibinfo{journal}{\emph{SIGPLAN Notices}} \bibinfo{volume}{17},
  \bibinfo{number}{9} (\bibinfo{date}{Sept.} \bibinfo{year}{1982}),
  \bibinfo{pages}{7–13}.
\newblock
\href{https://doi.org/10.1145/947955.1083808}{doi:\nolinkurl{10.1145/947955.1083808}}


\bibitem[Peyton~Jones(2003)]%
        {peytonjones2003wearing}
\bibfield{author}{\bibinfo{person}{Simon Peyton~Jones}.}
  \bibinfo{year}{2003}\natexlab{}.
\newblock \bibinfo{title}{Wearing the hair shirt: a retrospective on Haskell}.
  (\bibinfo{date}{January} \bibinfo{year}{2003}).
\newblock
\urldef\tempurl%
\url{https://www.microsoft.com/en-us/research/publication/wearing-hair-shirt-retrospective-haskell-2003/}
\showURL{%
\tempurl}
\newblock
\shownote{invited talk at POPL 2003}.


\bibitem[Raymond(2000)]%
        {raymond2000hackers}
\bibfield{author}{\bibinfo{person}{Eric Raymond}.}
  \bibinfo{year}{2000}\natexlab{}.
\newblock \showarticletitle{Guest Editorial: World Domination}.
\newblock \bibinfo{journal}{\emph{Linux Journal}} (\bibinfo{date}{Jan.}
  \bibinfo{year}{2000}).
\newblock
\urldef\tempurl%
\url{https://www.linuxjournal.com/article/3676}
\showURL{%
\tempurl}


\bibitem[Raymond(1990)]%
        {raymondintercal}
\bibfield{author}{\bibinfo{person}{Eric~S Raymond}.}
  \bibinfo{year}{1990}\natexlab{}.
\newblock \bibinfo{title}{INTERCAL}.
\newblock
\newblock
\shownote{\url{https://gitlab.com/esr/intercal}}.


\bibitem[Raymond(2010)]%
        {raymond2010risks}
\bibfield{author}{\bibinfo{person}{Eric~S Raymond}.}
  \bibinfo{year}{2010}\natexlab{}.
\newblock \bibinfo{title}{Risk, Verification, and the {INTERCAL} Reconstruction
  Massacree}.
\newblock
\newblock
\shownote{\url{http://esr.ibiblio.org/?p=2491}}.


\bibitem[Ritchie(1996)]%
        {ritchie1996development}
\bibfield{author}{\bibinfo{person}{Dennis~M. Ritchie}.}
  \bibinfo{year}{1996}\natexlab{}.
\newblock \showarticletitle{The development of the C programming language}. In
  \bibinfo{booktitle}{\emph{History of Programming Languages---II}}.
  \bibinfo{pages}{671–698}.
\newblock
\href{https://doi.org/10.1145/234286.1057834}{doi:\nolinkurl{10.1145/234286.1057834}}


\bibitem[Sammet and Hemmendinger(2003)]%
        {sammet2003programming}
\bibfield{author}{\bibinfo{person}{Jean~E. Sammet} {and} \bibinfo{person}{David
  Hemmendinger}.} \bibinfo{year}{2003}\natexlab{}.
\newblock \bibinfo{booktitle}{\emph{Programming languages}}.
\newblock \bibinfo{publisher}{John Wiley and Sons Ltd.},
  \bibinfo{pages}{1470–1475}.
\newblock
\showISBNx{0470864125}


\bibitem[Shaw(2022)]%
        {shaw2022myths}
\bibfield{author}{\bibinfo{person}{Mary Shaw}.}
  \bibinfo{year}{2022}\natexlab{}.
\newblock \showarticletitle{Myths and mythconceptions: what does it mean to be
  a programming language, anyhow?}
\newblock \bibinfo{journal}{\emph{Proc. ACM Program. Lang.}}
  \bibinfo{volume}{4}, \bibinfo{number}{HOPL}, Article \bibinfo{articleno}{234}
  (\bibinfo{date}{April} \bibinfo{year}{2022}).
\newblock
\href{https://doi.org/10.1145/3480947}{doi:\nolinkurl{10.1145/3480947}}


\bibitem[Steele and Gabriel(2008)]%
        {steele50in50}
\bibfield{author}{\bibinfo{person}{Guy Steele} {and} \bibinfo{person}{Richard
  Gabriel}.} \bibinfo{year}{2008}\natexlab{}.
\newblock \bibinfo{title}{50 in 50}.
\newblock
\newblock
\shownote{\url{http://lambda-the-ultimate.org/node/3101}}.


\bibitem[Steele(1999)]%
        {steele1999growing}
\bibfield{author}{\bibinfo{person}{Guy~L Steele}.}
  \bibinfo{year}{1999}\natexlab{}.
\newblock \showarticletitle{Growing a language}.
\newblock \bibinfo{journal}{\emph{Higher-order and symbolic computation}}
  \bibinfo{volume}{12}, \bibinfo{number}{3} (\bibinfo{year}{1999}),
  \bibinfo{pages}{221--236}.
\newblock
\href{https://doi.org/10.1023/A:1010085415024}{doi:\nolinkurl{10.1023/A:1010085415024}}


\bibitem[Stroustrup(1998)]%
        {stroustrup1998c}
\bibfield{author}{\bibinfo{person}{Bjarne Stroustrup}.}
  \bibinfo{year}{1998}\natexlab{}.
\newblock \bibinfo{title}{Generalizing Overloading for {C}++2000}.
\newblock
\newblock
\shownote{\url{https://www.stroustrup.com/whitespace98.pdf}}.


\bibitem[Swidan and Hermans(2023)]%
        {swidan2023framework}
\bibfield{author}{\bibinfo{person}{Alaaeddin Swidan} {and}
  \bibinfo{person}{Felienne Hermans}.} \bibinfo{year}{2023}\natexlab{}.
\newblock \showarticletitle{A Framework for the Localization of Programming
  Languages}. In \bibinfo{booktitle}{\emph{Proceedings of the 2023 ACM SIGPLAN
  International Symposium on SPLASH-E}}. \bibinfo{pages}{13–25}.
\newblock
\href{https://doi.org/10.1145/3622780.3623645}{doi:\nolinkurl{10.1145/3622780.3623645}}


\bibitem[Syme(2020)]%
        {syme2020early}
\bibfield{author}{\bibinfo{person}{Don Syme}.} \bibinfo{year}{2020}\natexlab{}.
\newblock \showarticletitle{The early history of F\#}.
\newblock \bibinfo{journal}{\emph{Proc. ACM Program. Lang.}}
  \bibinfo{volume}{4}, \bibinfo{number}{HOPL}, Article \bibinfo{articleno}{75}
  (\bibinfo{date}{June} \bibinfo{year}{2020}), \bibinfo{numpages}{58}~pages.
\newblock
\href{https://doi.org/10.1145/3386325}{doi:\nolinkurl{10.1145/3386325}}


\bibitem[Temkin(2009)]%
        {velato}
\bibfield{author}{\bibinfo{person}{Daniel Temkin}.}
  \bibinfo{year}{2009}\natexlab{}.
\newblock \bibinfo{title}{Velato}.
\newblock
\newblock
\shownote{\url{http://velato.net}}.


\bibitem[Temkin(2017)]%
        {temkin2017language}
\bibfield{author}{\bibinfo{person}{Daniel Temkin}.}
  \bibinfo{year}{2017}\natexlab{}.
\newblock \showarticletitle{Language without code: intentionally unusable,
  uncomputable, or conceptual programming languages}.
\newblock \bibinfo{journal}{\emph{Journal of Science and Technology of the
  Arts}} \bibinfo{volume}{9}, \bibinfo{number}{3} (\bibinfo{date}{Sep.}
  \bibinfo{year}{2017}), \bibinfo{pages}{83--91}.
\newblock
\href{https://doi.org/10.7559/citarj.v9i3.432}{doi:\nolinkurl{10.7559/citarj.v9i3.432}}


\bibitem[Temkin(2023)]%
        {temkin2023less}
\bibfield{author}{\bibinfo{person}{Daniel Temkin}.}
  \bibinfo{year}{2023}\natexlab{}.
\newblock \showarticletitle{The Less Humble Programmer}.
\newblock \bibinfo{journal}{\emph{DHQ: Digital Humanities Quarterly}}
  \bibinfo{volume}{17}, \bibinfo{number}{2} (\bibinfo{year}{2023}).
\newblock


\bibitem[Temkin(2025)]%
        {temkin2025esolang}
\bibfield{author}{\bibinfo{person}{Daniel Temkin}.}
  \bibinfo{year}{2025}\natexlab{}.
\newblock \bibinfo{booktitle}{\emph{Forty-Four Esolangs: The Art of Esoteric
  Code}}.
\newblock \bibinfo{publisher}{MIT Press}.
\newblock
\showISBNx{9780262553087}
\newblock
\shownote{(to appear)}.


\bibitem[Thuné and and(2009)]%
        {thune2009variation}
\bibfield{author}{\bibinfo{person}{Michael Thuné} {and}
  \bibinfo{person}{Anna~Eckerdal and}.} \bibinfo{year}{2009}\natexlab{}.
\newblock \showarticletitle{Variation theory applied to students’ conceptions
  of computer programming}.
\newblock \bibinfo{journal}{\emph{European Journal of Engineering Education}}
  \bibinfo{volume}{34}, \bibinfo{number}{4} (\bibinfo{year}{2009}),
  \bibinfo{pages}{339--347}.
\newblock
\href{https://doi.org/10.1080/03043790902989374}{doi:\nolinkurl{10.1080/03043790902989374}}


\bibitem[Tiwari et~al\mbox{.}(2024)]%
        {tiwari2024great}
\bibfield{author}{\bibinfo{person}{Deepika Tiwari}, \bibinfo{person}{Tim
  Toady}, \bibinfo{person}{Martin Monperrus}, {and} \bibinfo{person}{Benoit
  Baudry}.} \bibinfo{year}{2024}\natexlab{}.
\newblock \showarticletitle{With Great Humor Comes Great Developer Engagement}.
  In \bibinfo{booktitle}{\emph{Proceedings of the 46th International Conference
  on Software Engineering: Software Engineering in Society}}.
  \bibinfo{pages}{1–11}.
\newblock
\href{https://doi.org/10.1145/3639475.3640099}{doi:\nolinkurl{10.1145/3639475.3640099}}


\bibitem[Tshukudu and Cutts(2020)]%
        {tshukudu2020understanding}
\bibfield{author}{\bibinfo{person}{Ethel Tshukudu} {and}
  \bibinfo{person}{Quintin Cutts}.} \bibinfo{year}{2020}\natexlab{}.
\newblock \showarticletitle{Understanding conceptual transfer for students
  learning new programming languages}. In \bibinfo{booktitle}{\emph{Proceedings
  of the 2020 ACM conference on international computing education research}}.
  \bibinfo{pages}{227--237}.
\newblock
\href{https://doi.org/10.1145/3372782.3406270}{doi:\nolinkurl{10.1145/3372782.3406270}}


\bibitem[Twist et~al\mbox{.}(2025)]%
        {twist2025llms}
\bibfield{author}{\bibinfo{person}{Lukas Twist}, \bibinfo{person}{Jie~M.
  Zhang}, \bibinfo{person}{Mark Harman}, \bibinfo{person}{Don Syme},
  \bibinfo{person}{Joost Noppen}, {and} \bibinfo{person}{Detlef Nauck}.}
  \bibinfo{year}{2025}\natexlab{}.
\newblock \bibinfo{title}{LLMs Love Python: A Study of LLMs' Bias for
  Programming Languages and Libraries}.
\newblock
\showeprint[arxiv]{2503.17181}~[cs.SE]
\urldef\tempurl%
\url{https://arxiv.org/abs/2503.17181}
\showURL{%
\tempurl}


\bibitem[Wang et~al\mbox{.}(2024)]%
        {wang2024exploring}
\bibfield{author}{\bibinfo{person}{Chaozheng Wang}, \bibinfo{person}{Zongjie
  Li}, \bibinfo{person}{Cuiyun Gao}, \bibinfo{person}{Wenxuan Wang},
  \bibinfo{person}{Ting Peng}, \bibinfo{person}{Hailiang Huang},
  \bibinfo{person}{Yuetang Deng}, \bibinfo{person}{Shuai Wang}, {and}
  \bibinfo{person}{Michael~R. Lyu}.} \bibinfo{year}{2024}\natexlab{}.
\newblock \bibinfo{title}{Exploring Multi-Lingual Bias of Large Code Models in
  Code Generation}.
\newblock
\showeprint[arxiv]{2404.19368}~[cs.SE]
\urldef\tempurl%
\url{https://arxiv.org/abs/2404.19368}
\showURL{%
\tempurl}


\bibitem[Wing(2006)]%
        {wing2006computational}
\bibfield{author}{\bibinfo{person}{Jeannette~M Wing}.}
  \bibinfo{year}{2006}\natexlab{}.
\newblock \showarticletitle{Computational thinking}.
\newblock \bibinfo{journal}{\emph{Commun. ACM}} \bibinfo{volume}{49},
  \bibinfo{number}{3} (\bibinfo{year}{2006}), \bibinfo{pages}{33--35}.
\newblock
\href{https://doi.org/10.1145/1118178.1118215}{doi:\nolinkurl{10.1145/1118178.1118215}}


\bibitem[Woods and Lyon(1973)]%
        {intercalman}
\bibfield{author}{\bibinfo{person}{Donald~R. Woods} {and}
  \bibinfo{person}{James~M. Lyon}.} \bibinfo{year}{1973}\natexlab{}.
\newblock \bibinfo{title}{The {INTERCAL} Programming Language Reference
  Manual}.
\newblock
\newblock
\shownote{\url{https://3e8.org/pub/intercal.pdf}}.


\bibitem[Yuan et~al\mbox{.}(2024)]%
        {please}
\bibfield{author}{\bibinfo{person}{Yicong Yuan}, \bibinfo{person}{Mingyang Su},
  {and} \bibinfo{person}{Xiu Li}.} \bibinfo{year}{2024}\natexlab{}.
\newblock \showarticletitle{What Makes People Say Thanks to {AI}}. In
  \bibinfo{booktitle}{\emph{Artificial Intelligence in {HCI}}},
  \bibfield{editor}{\bibinfo{person}{Helmut Degen} {and}
  \bibinfo{person}{Stavroula Ntoa}} (Eds.). \bibinfo{publisher}{Springer Nature
  Switzerland}, \bibinfo{address}{Cham}, \bibinfo{pages}{131--149}.
\newblock
\showISBNx{978-3-031-60606-9}
\href{https://doi.org/10.1007/978-3-031-60606-9_9}{doi:\nolinkurl{10.1007/978-3-031-60606-9_9}}


\end{thebibliography}
%%%%%%%%%%%%%%%%%%%%
\end{document}